\documentclass[12pt,draftcls,onecolumn]{IEEEtran}
\usepackage{amsmath,graphicx,float,cite,dsfont,amssymb}
\usepackage{amsthm,amsfonts,multirow,bm,array,setspace,stfloats}
\usepackage{amsmath}
\usepackage{cite}
\usepackage{mathtools}
\newcommand{\defeq}{\vcentcolon=}
\newcommand{\eqdef}{=\vcentcolon}
\usepackage{textcomp}
\usepackage{psfrag}
\usepackage{color}
\usepackage{pgfplots}
\usepackage{graphicx}
\usepackage{listings}
\usepackage{algorithm}
\usepackage{algcompatible}
\usepackage{tikz}
\usetikzlibrary{arrows}
\usetikzlibrary{patterns}
\tikzset{>=latex'}
\usetikzlibrary{shapes.geometric}
\usepackage{multirow}
\usepackage{graphicx,subfigure}
\usetikzlibrary{calc}
\usetikzlibrary{positioning, shapes}

\usepackage{mathtools}
\usepackage[skip=2pt,font=small]{caption}
\usepackage{tabularx,ragged2e,booktabs,caption}
\usepackage{balance}
\usepackage{xpatch,letltxmacro}
\makeatletter
\LetLtxMacro\oldalgorithmic\algorithmic
\LetLtxMacro\endoldalgorithmic\endalgorithmic
\renewenvironment{algorithmic}[1][0]{%
	\oldalgorithmic[#1]%
	\xpatchcmd{\REPEAT}{\ALC@it}{\ALC@it@nostep}{}{}%
	\xpatchcmd{\UNTIL}{\ALC@it}{\ALC@it@nostep}{}{}%
}{\endoldalgorithmic}
\makeatother
\makeatletter
\def\endthebibliography{%
  \def\@noitemerr{\@latex@warning{Empty `thebibliography' environment}}%
  \endlist
}
\usepackage[utf8]{inputenc}
\usepackage[english]{babel}

\usepackage{amsthm}
\DeclareMathOperator{\EX}{\mathbb{E}}
\DeclarePairedDelimiter\normV{\lVert}{\rVert}

\newcommand\raisepunct[1]{\,\mathpunct{\raisebox{0.5ex}{#1}}}

\theoremstyle{remark}
\newtheorem{theorem}{ {Theorem}}

\newtheorem{proposition}{{Proposition}}

\let\emptyset\varnothing

\def\hexagonsize{1cm}
\pgfdeclarepatternformonly
{hexagons}
{\pgfpointorigin}
{\pgfpoint{3*\hexagonsize}{0.866025*2*\hexagonsize}}
{\pgfpoint{3*\hexagonsize}{0.866025*2*\hexagonsize}}
{
	\pgfsetlinewidth{0.4pt}
	\pgftransformshift{\pgfpoint{0mm}{0.866025*\hexagonsize}}
	\pgfpathmoveto{\pgfpoint{0mm}{0mm}}
	\pgfpathlineto{\pgfpoint{0.5*\hexagonsize}{0mm}}
	\pgfpathlineto{\pgfpoint{\hexagonsize}{-0.866025*\hexagonsize}}
	\pgfpathlineto{\pgfpoint{2*\hexagonsize}{-0.866025*\hexagonsize}}
	\pgfpathlineto{\pgfpoint{2.5*\hexagonsize}{0mm}}
	\pgfpathlineto{\pgfpoint{3*\hexagonsize+0.2mm}{0mm}}
	\pgfpathmoveto{\pgfpoint{0.5*\hexagonsize}{0mm}}
	\pgfpathlineto{\pgfpoint{\hexagonsize}{0.866025*\hexagonsize}}
	\pgfpathlineto{\pgfpoint{2*\hexagonsize}{0.866025*\hexagonsize}}
	\pgfpathlineto{\pgfpoint{2.5*\hexagonsize}{0mm}}
	\pgfusepath{stroke}
}

\ifCLASSINFOpdf
\else
\fi
\begin{document}
%
\title{Rate Splitting Multiple Access in C-RAN: A Scalable and Robust Design}
\author{\IEEEauthorblockN{Alaa Alameer Ahmad, \IEEEmembership{Student Member, IEEE}, Yijie Mao, \IEEEmembership{Member, IEEE}, Aydin Sezgin, \IEEEmembership{Senior Member, IEEE}, and Bruno Clerckx, \IEEEmembership{Senior Member, IEEE}\\}
\thanks{Part of this paper was presented at the IEEE International Symposium on Personal, Indoor and Mobile Radio Communications (PIMRC), virtual conference, September 2020 \cite{9217249}.\newline \indent This work has been partially supported by the German Research Foundation, Deutsche Forschungsgemeinschaft (DFG), Germany, under the project ATINA with grant number SE1697/19-1 and by the U.K. Engineering and Physical Sciences Research Council (EPSRC) under grant EP/N015312/1, EP/R511547/1. \newline
\IEEEauthorblockA{\indent Alaa Alameer Ahmad and Aydin Sezgin are with Digital Communication Systems, Ruhr-Universit\"at Bochum, Bochum, Germany. (Email: \{alaa.alameerahmad,aydin.sezgin\}@rub.de)}\newline
\indent Yijie Mao and Bruno Clerckx are with with the Communications and Signal Processing group, Department of Electrical and Electronic Engineering, Imperial College London, U.K. (Email: \{y.mao16, b.clerckx\}@imperial.ac.uk)
}
}
\maketitle
\vspace{-1.2cm}
\begin{abstract}
Cloud radio access networks (C-RAN) enable a network platform for beyond the fifth generation of communication networks (B5G), which incorporates the advances in cloud computing technologies to modern radio access networks. Recently, rate splitting multiple access (RSMA), relying on multi-antenna rate-splitting (RS) at the transmitter and successive interference cancellation (SIC) at the receivers, has been shown to manage the interference in multi-antenna communication networks efficiently. This paper considers applying RSMA in C-RAN. We address the practical challenge of a transmitter that only knows the statistical channel state (CSI) information of the users. To this end, the paper investigates the problem of stochastic coordinated beamforming (SCB) optimization to maximize the ergodic sum-rate (ESR) in the network. Furthermore, we propose a scalable and robust RS scheme where the number of the common streams to be decoded at each user scales linearly with the number of users, and the common stream selection only depends on the statistical CSI. The setup leads to a challenging stochastic and non-convex optimization problem. A sample average approximation (SAA) and weighted minimum mean square error (WMMSE) based algorithm is adopted to tackle the intractable stochastic non-convex optimization and guarantee convergence to a stationary point asymptotically. The numerical simulations demonstrate the efficiency of the proposed RS strategy and show a gain up to 27\% in the achievable ESR compared with state-of-the-art schemes, namely treating interference as noise (TIN) and non-orthogonal multiple access (NOMA) schemes.
\end{abstract}
\section{Introduction}
Current wireless networks are forecast to face a tremendous increase in the number of connected people and devices in the next years. Thus, the B5G of mobile wireless networks is expected to handle a massive amount of data-traffic mainly driven by applications providing Video-on-Demand such as Youtube and Netflix \cite{DBLP,8782879}. C-RAN allows B5G networks to cope with the huge increase in data traffic by enabling spatial reuse through network densification which improves the performance of wireless networks. In C-RAN, a central processor (CP) utilizes cloud computing techniques to manage a set of base stations (BSs) distributed within the network. The BSs are connected to the CP via high-speed and limited capacity fronthaul links. In this network architecture, most of the baseband signal processing tasks are migrated to the CP, which centrally controls users' data, enables fully cooperative transmission and resource allocation, and can achieve the best performance by exploiting the benefits of large-scale virtual multiple-input multiple-output (MIMO) systems \cite{1678166}.

C-RAN can significantly boost the spectrum efficiency (SE) and energy efficiency (EE) \cite{97813165296}, thanks to its ability to support efficient cooperative transmission schemes such as coordinated beamforming. However, to exploit the full-scale gain of these cooperative schemes and to optimize the transmission strategy (e.g., linear beamforming design), most algorithms in the literature require perfect knowledge of channel state information (CSI) at the CP \cite{6920005, 6786060}.

The assumption of perfect CSI availability is often rather optimistic and impractical. In practice, it is very hard to perfectly estimate the channel coefficients at the CP for each transmits operation, and the uncertainty of CSI is inevitable. The source of CSI uncertainties can be due to channel estimation errors in time-division duplex (TDD) systems \cite{1193803} or quantization errors in limited feedback frequency division duplex (FDD) systems \cite{4641946}. Other sources of imperfections may arise from hardware limitations, channel estimation delays, and channel aging. Moreover, in dense networks where many BSs are participating in transmission, obtaining the full CSI of all links involves a significant training overhead that is infeasible in practical scenarios. 
The channel knowledge uncertainty brings new challenges to the system design. Hence, the problem of ergodic sum-rate maximization when considering imperfect channel state information at the transmitter (CSIT) is an optimization problem of stochastic nature and requires different optimization tools than those usually used to approach the deterministic counterpart problems when full CSIT is assumed. In the literature, two methods exist to deal with this kind of stochastic optimization problems. One approach is to use a robust beamforming vector optimization framework. In this line of research, the stochastic optimization problem is transformed into its deterministic counterpart. Specifically, in robust optimization, the channel uncertainty model is deterministic, and it is characterized based on set models \cite{BenRob}, which makes robust optimization tractable from a computational point of view. However, the robust optimization model is often over-conservative \cite{10.2307/23070141} and is solely designed to account for the worst-case scenario, which usually happens with very low probability \cite{5447076}.

The second approach tackles the stochastic problem in an alternative way where the aim is to design the beamforming vectors such that the performance of the system is optimized on average, which yields better performance than the conservative design of robust optimization framework. This approach uses stochastic optimization techniques \cite{StochSh} to design the beamforming vectors and maximize the system's performance, based on statistical CSIT knowledge which is easy to obtain in practice \cite{8664604} rather than requiring the full CSIT knowledge, which is very hard to obtain in dense networks such as C-RAN. Motivated by this fact and the performance of the stochastic optimization approach as compared with other conservative approaches, we propose in this paper to use a stochastic programming framework to deal with the uncertainty in the CSI. We do not require any assumptions on the channel uncertainty models, but we assume only the channel distribution knowledge, which is reasonable in C-RAN.

An interesting observation is that, on the one hand, C-RAN can be viewed as a broadcast channel (BC) when the fronthaul capacity tends to infinity as it becomes equivalent to a virtual MIMO system. While on the other hand, it can be interpreted as an interference channel (IC) in the zero fronthaul capacity extreme as the cooperation among BSs becomes infeasible. That is, C-RAN generalizes and bridges both setups. The capacity of the IC channel is a well-known open problem in information theory. Rate splitting (RS) and common message decoding (CMD) is to date the best-known transmission scheme for the IC \cite{1056307,4675741}.
RS dates back to the early '80s in the work of \cite{Carleial78}. In RS, the transmitter first splits each message into private and common parts and then encodes the private parts into private streams, and the common parts are encoded into common streams. The private streams are decoded by the respective receivers only, while a group of receivers decodes the common streams to mitigate the interference within this group. With RS, the presence of common messages adds extra flexibility in terms of rates allocation, beamforming vectors design, and decoding options among the users. Such flexibility helps to reduce the interference and provides a significant  performance improvement as compared to other transmission schemes which do not use common messages such as TIN. Motivated by the information-theoretical insights, this paper considers RS transmission strategies in a C-RAN downlink. The CP applies central joint encoding to private and common parts of all users' messages and enables the cooperation between the BSs by the joint design of beamforming vectors. In particular, the CP shares the private and common streams with cooperating clusters of BSs. This kind of data exchange between the CP and BSs through the fronthaul links is referred to in the literature as \textit{data-sharing}.

On the one hand, most works in C-RAN literature focus on full CSIT assumption with TIN strategy. On the other hand, RS is extensively studied in the special cases of IC and multi-antenna BC-BS and shows robustness against CSIT imperfections compared with TIN. Motivated by this fact, this work considers a realistic scenario of imperfect CSIT at the CP and investigates an RS-enabled C-RAN system. We consider the problem of maximizing the ergodic sum-rate, subject to per-BS transmit power and fronthaul capacity constraints. Thus, the problem aims at finding the optimally coordinated beamforming vectors design and rate allocation of streams such that the ESR performance in the network is maximized in the absence of full CSIT.
\subsection{Related Work}
The problem of sum-rate (SR) maximization in C-RAN with full CSI assumption is extensively studied \cite{6588350, 6920005,Simeone2009,7105959}. However, in practice, the assumption of full CSI is unrealistic, especially in dense networks. The authors in \cite{7185432, 8606433, 7315049} have studied the SR problem with imperfect CSI. In particular, the authors in \cite{7185432} consider the SR under ergodic CSI and limited fronthaul capacity assumptions. The resulting ergodic SR problem is then solved by merging the majorization–minimization (MM) \cite{6588350} and the stochastic successive upper bound minimization (SSUM) framework \cite{SSUM}. The work in \cite{8606433} considers imperfect CSI and proposes a global optimization algorithm and a reduced-complexity modified weighted minimum mean square error (WMMSE) algorithm to solve a relaxed ergodic SR problem where a rate lower bound is maximized.

Other works consider imperfect CSI in C-RAN and focus on minimizing the power consumption in the network \cite{6996028}, or investigate a multi-objective performance criterion such as the energy efficiency in the network \cite{8003504}. All these works \cite{6588350, 6996028, 8003504, 6920005,Simeone2009,7105959,7185432, 8606433, 7315049} consider the baseline TIN as a communication scheme in which the CP encodes private streams (of unicast nature) decoded by intended users only, and each user treats the interference of other users as noise.

As aforementioned, from an information-theoretical point of view, the TIN strategy is, in general, suboptimal and can degrade the performance of the system significantly, especially in interference-limited dense networks \cite{7322253, 6123786}. In \cite{1056307} the authors show that RS and common message decoding (CMD) scheme can achieve the capacity of a 2-user interference channel (IC) in special cases. In general, the capacity of the IC is still an open problem. However, in the seminal work \cite{4675741}, the authors show that RS-CMD can achieve within one-bit of the IC's capacity.

Inspired by these information-theoretical works \cite{Carleial78, 1056307, 4675741}, which study a basic 2-user IC, the authors in \cite{7422790} show that the RS strategy can approximate the capacity of the symmetric $K$-user Gaussian IC with substantial interference to within a constant gap. The authors of \cite{5910112} apply RS-CMD in a practical multi-cell communication scenario and show the gain of RS compared to TIN in such a network.

Several works have recently studied RS techniques in multiple input single output broadcast channel (MISO-BC) \cite{7513415,7434643,7555358,8907421,7470942, Mao2, Mao3, Mao4 ,Mao,9120024} as a promising transmission strategy candidate to manage the interference in B5G networks. The key benefit of RS in multi-antenna BC is the information-theoretical optimality in the sense of achieving the optimal degrees of freedom (DoF) region of the MISO BC with imperfect CSIT \cite{7972900}. This DoF optimality translates into significant rate gains, such that RS was shown to outperform Dirty Paper Coding (DPC) in \cite{mao2019dirty}.

As opposed to the simple 2-user IC, in MISO-BC, RS design needs to specify the group of users who decode each common stream (of multicast nature), which is a non-trivial task. The authors in \cite{7555358} propose a simple RS scheme which consists of a common stream with a broadcast nature, i.e., it is decoded by all users while private streams are decoded by intended users only. The authors in \cite{7555358} study the problem of SR maximization with partial CSIT and show that such a simple RS strategy achieves considerable gains compared to TIN, especially in strong SNR regimes.
In \cite{Mao}, the authors propose a more general flexible RSMA, based on a generalized RS scheme, in which the message-splitting and combining is done hierarchically. The generalized RS is shown to generalize both extremes, TIN, and non-orthogonal multiple access (NOMA). However, the gain of generalized RS comes at an exponential complexity in the number of users when message splitting is performed.

All the works in \cite{7513415,7434643,7555358,8907421,7470942, Mao2, Mao3, Mao4 ,Mao,9120024,7972900,mao2019dirty} investigate RS techniques in a MISO-BC network. As mentioned previously, the MISO-BC is a special case of C-RAN with unlimited fronthaul capacity. Thus, it is interesting to investigate the interplay between the limited fronthaul capacity, cooperative transmission schemes, and RS techniques. Moreover, with its network architecture, C-RAN can handle RS techniques very efficiently, and the implementation of RS algorithms can be smoothly realized, with the help of cloud computing technologies. In this context, the authors in \cite{8756076} investigate generalized RS in a C-RAN, which adopts a compression strategy in the downlink. In \cite{8732995} the authors propose a simple RS scheme in which the message of each user is split into a private message to be decoded by the intended user and a common message to be decoded by a group of users. The procedure in \cite{8732995}, is to determine the set of users decoding the common message of each user. The authors in \cite{8732995} propose a polynomial complexity algorithm to jointly optimize the beamforming vectors, rate allocation and BSs clusters design to maximize the SR in a downlink C-RAN with data-sharing. Although the RS scheme in \cite{8732995} is much simpler than the generalized RS, it requires full CSIT at the CP. Both works \cite{8756076,8732995} assume full CSIT at the CP. However, CSIT uncertainties introduce technical challenges to system design and performance optimization. Thus, the algorithm developed in \cite{8756076,8732995} can not be applied if the full CSIT assumption is not fulfilled. Motivated by the benefits of RS in elastic management of the interference in communication networks, the paper at hand employs RS techniques in C-RAN, where the CP has only statistical information on the wireless channel. Next, we discuss the main contributions of this paper.
\subsection{Contributions}
In this paper, we investigate employing RS techniques in C-RAN with statistical CSI at the CP. To the best of our knowledge, this is the first paper that addresses the generalized RS scheme in C-RAN with CSIT imperfections. Although the generalized RS is robust against CSIT imperfections, it suffers from the curse of scalability. Hence, the number of common streams in generalized RS scales exponentially with the number of users, which leads to being impractical to use generalized RS in dense networks. Therefore, we further propose a practical RS scheme that is robust against CSI imperfections and scalable with the number of users.
The major contributions are as follows
\begin{itemize}
\item[1)] {\textit{Novel practical RS scheme:}} We propose a practical RS scheme that enjoys the robustness against CSI imperfections and the scalability with the number of users. In contrast to the approach in \cite{8732995}, the design of the proposed RS scheme requires the knowledge of statistical CSI only, which can be efficiently and accurately acquired as opposed to the full CSIT requirement. Furthermore, the number of common streams in the proposed RS scheme scales linearly with the number of users as opposed to the generalized RS scheme \cite{Mao}. We mention here that the one-layer RS scheme proposed in \cite{7555358} is also scalable as it only uses one common message irrespectively of the number of users and does not require full CSIT. However, we show in the numerical simulations that our proposed RS scheme significantly outperforms the performance of the one-layer RS, especially in dense networks. Interestingly, the proposed scheme performs close to the generalized RS scheme, which involves a considerably higher number of common messages.


\item[2)] \textit{Optimization framework:} We formulate a joint rate allocation and stochastic coordinated beamforming (SCB) optimization to maximize the ergodic sum-rate of the wireless network subject to per-BS transmit power and fronthaul capacity constraints. In \cite{7555358}, the authors address an ESR maximization problem in MISO-BC. However, MISO-BC can be viewed as a special case of C-RAN in which the fronthaul capacity is infinite, and the transmitter antennas are co-located (no BSs clustering is needed). In contrast to \cite{7555358}, this paper explicitly accounts for the limited capacity of fronthaul links and the cooperative beamforming design in C-RAN. Hence, the optimization approaches of \cite{8732995, 7555358}, can not be directly applied to our problem. Therefore, we propose a general solution to the problem, which merges the sample average approximation method and WMMSE-rate relationship and develops a block coordinated assent (BCA) iterative algorithm. Our proposed solution is guaranteed to converge to a Karush-Kuhn-Tucker (KKT) point. The developed algorithm also applies to the generalized RS scheme, which extends the RS strategy considered in \cite{7555358}.
\item[3)] \textit{Clustering:} To account for message sharing between the cloud and the BSs, we propose a stream-based BSs clustering procedure, which requires only the statistical CSI and therefore it is of particular interest for practical scenarios. As opposed to the user-centric based static clustering algorithm in \cite{6920005}, we propose a stream-based clustering to account for the multicasting nature of common messages and the unicasting nature of private streams.
\item[4)] \textit{Numerical Simulations:} We perform extensive numerical simulations to evaluate the performance of the proposed scheme against the TIN, NOMA, and generalized RS schemes in a practical C-RAN system. In particular, we show the gain of the proposed algorithm in different practical scenarios.
\end{itemize}
Next, we give an overview of the organization of the paper.
\subsection{Organization}
The remainder of the paper is organized as follows: Section \ref{sec2} presents the system model and the rate splitting transmission schemes. Section \ref{sec3} introduces the signal model, the achievable rates, and the stochastic optimization problem formulation. The optimization framework is then introduced in section \ref{sec4}, which presents the SAA based approach coupled with the WMMSE algorithm. In section \ref{NS}, the numerical simulations are introduced, and finally, in section \ref{Con}, the paper is concluded and summarized. 
\begin{figure}
	\begin{center}
		\includegraphics[width=0.5\linewidth]{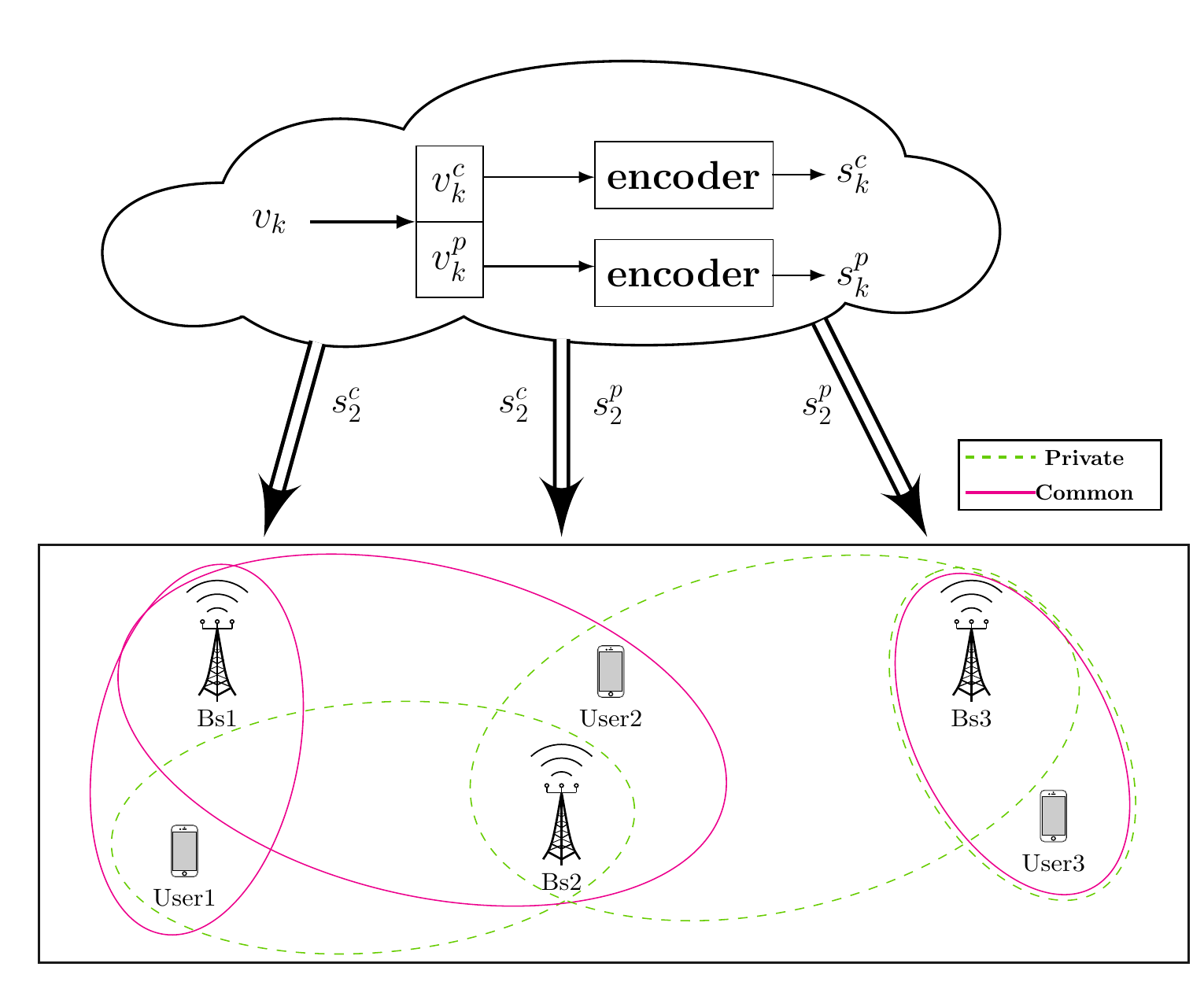}
		\caption{A C-RAN system with three BSs serving three users. Private and common messages are encoded at the cloud. As a simple illustration, we show the data sharing of User 2 with the BSs.}
	\end{center}
   \vspace{-0.5cm}
\end{figure}

\section{System Model and Rate Splitting}\label{sec2}
We consider a C-RAN system operating in downlink mode with a transmission bandwidth $B$. The network consists of a set of multiple-antenna base stations (BSs) $\mathcal{N} = \left\lbrace1,2,\ldots,N\right\rbrace$, serving a group of single-antenna users $\mathcal{K} = \left\lbrace1,2,\ldots,K\right\rbrace$. Each BS $n \in \mathcal{N}$ is equipped with $L > 1$ antennas and is connected to a central processor (CP), located at the cloud, via a fronthaul link of capacity $F_n$.
The received signal at user $k$ according to this scenario is
\begin{equation} \label{e0}
y_k = \mathbf{h}_k^{\dagger} \mathbf{x} + n_k.
\end{equation}
where $\left( \cdot\right)^T$ is the transpose operator, $\mathbf{h}_k  = \left[\mathbf{h}_{1,k}^T,\ldots, \mathbf{h}_{N,k}^T \right]^T \in \mathbb{C}^{NL\times 1}$ is the aggregate channel vector of user $k$ to all BSs in the network, $\mathbf{x}  = \left[\mathbf{x}_1^T,\ldots, \mathbf{x}_N^T\right] \in \mathbb{C}^{NL\times 1}$ is the aggregate transmit signal from all BSs, and
$n_k \sim \mathcal{CN}\left(0, \sigma^2\right)$ is the additive white Gaussian noise (AWGN).
The transmit signal $\mathbf{x}_n$ from BS $n$, is subject to the following per-BS maximum transmit power constraint $\EX\left\lbrace \mathbf{x}_n^{\dagger} \mathbf{x}_n\right\rbrace \leq {P}_n^{\text{Max}} $. This paper considers an ergodic block-fading channel model. The channel remains constant for a coherence period, which spans a block of channel uses and varies randomly in an ergodic fashion from one block to another. Next, we discuss the channel model adopted in this paper.
\subsection{Stochastic CSI Model}
We assume the CP has only information about the distribution of channel states, e.g., statistical CSI. However, receivers are considered to have perfect knowledge of the channel states, i.e., we assume perfect CSIR is available at users so that user $k$ knows the channel state $\mathbf{h}_k$ perfectly. These assumptions are quite general and can model different inaccuracies in CSI.
Considering the imperfect CSIT model, the channel between BS $n$ and user $k$ follows the distribution $\mathbf{h}_{n, k} \sim \mathbf{\mathcal{CN}}\left(0, \mathbf{Q}_{n, k}\right)$, where $\mathbf{Q}_{n, k}$ is a symmetric positive semi-definite matrix and represents the channel covariance matrix which depends on the path-loss between BS $n$ and user $k$. In particular, this assumption is practical because the path-loss information varies slowly and needs to be updated solely when the users' location changes which significantly reduces the overhead due to the CSIT acquisition process at the CP.
In what follows, by imperfect CSI, we mean that the CP is only aware of the covariance matrices, while full CSI implies that the CP has full knowledge of the coefficients $\left\lbrace \mathbf{h}_{n,k}| \forall k \in \mathcal{K}, \forall n \in \mathcal{N}\right\rbrace $.
 \subsection{Rate Splitting and Common Message Decoding with Data-Sharing}
In the literature, there exist two main methods to apply RS techniques in practical communication networks. Specifically, the first approach is
based on rate-splitting and common message decoding (RS-CMD) as proposed in \cite{8732995}, and the second one is the generalized RS scheme as introduced in \cite{Mao}. In this work, we focus on the RS-CMD scheme and use the generalized RS and sub-schemes thereof as benchmarking schemes.

Specifically, in RS-CMD with data sharing,
the message of each user $k \in \mathcal{K}$, $v_k$ is split into two messages only, a private message (of unicast nature) $v_{k}^p$ decoded by the intended user and a common message (of multicast nature) $v_{k}^c$ decodable by a group of users.
The private and common messages are then encoded at the CP to produce the coded streams $s_k^p$ and $s_k^c$, respectively. As opposed to MISO-BC in which a single transmitter sends messages to a set of users, in C-RAN, the CP shares each coded stream of each user with a subset of BSs, which cooperatively transmit the beamformed signal to the corresponding users. Let $\mathcal{K}_n^p,\mathcal{K}_n^c \subseteq \mathcal K$ be the subset of users served by BS $n$ with a private or common message, respectively, i.e.,
\begin{align}
\mathcal{K}_{n}^{p} &\defeq \left\lbrace k \in \mathcal{K}|\hspace{1mm} \text{BS \textit{n} delivers}\hspace{1mm} s_{k}^p \hspace{1mm} \text{to user}\hspace{1mm} k\right\rbrace,\\
\mathcal{K}_{n}^{c} &\defeq \left\lbrace k \in \mathcal{K}|\hspace{1mm} \text{BS \textit{n} delivers}\hspace{0.5mm} s_{k}^c \hspace{1mm} \text{to user}\hspace{1mm} k\right\rbrace.
\end{align}
Upon receiving the encoded symbols, BS $n$ constructs the transmit signal $\mathbf{x}_n \in \mathbb{C}^{L \times 1}$ by a superposition of linearly precoded streams with beamforming vectors $\mathbf{w}_{n, k}^{p}, \mathbf{w}_{n, k}^{c} \in \mathbb{C}^{L \times 1}$ as follows \vspace{-0.2cm}
\begin{equation}
\mathbf{x}_n = \sum_{k \in \mathcal{K}_{n}^p} \mathbf{w}_{n, k}^{p} s_{k}^{p} + \sum_{k \in \mathcal{K}_{n}^c} \mathbf{w}_{n, k}^{c} s_{k}^{c}.
\vspace{-0.2cm}
\end{equation}
Note that since the fronthaul capacity is finite, each BS $n$ can only support a limited number of streams. If BS $n$ does not participate in transmitting stream $s_{k}^p$, then $\mathbf{w}_{n, k}^{p} = \mathbf{0}_L$, where $\mathbf{0}_L$ is an all-zero vector of length $L$. Similarly if $s_{k}^c$ is not served by BS $n$ then $\mathbf{w}_{n, k}^{c} = \mathbf{0}_L$. 

The main challenge in applying the RS-CMD scheme in practical communication networks such as C-RAN is to determine the group of users that decode each user's common message. Let $\mathcal{M}_k$ denote the common message set be defined as \vspace{-0.2cm}
\begin{equation}
\mathcal{M}_k = \left\lbrace j \in \mathcal{K}| \hspace{1mm}\text{user} \hspace{1mm}j\hspace{1mm}  \text{decodes the common message of user} \hspace{1mm}k \right\rbrace.
\vspace{-0.2cm}
\end{equation}
In this work, we propose a practical procedure to determine $\mathcal{M}_k$, which does not require channel knowledge. Hence, it is robust against CSI imperfections. This method is solely based on the geographical locations of the users. This information can be easily obtained using global positioning systems (GPS) devices \cite{8664604}.

In particular, let $d_{k_1 k_2}$ denotes the distance between users $k_1$ and $k_2$ and $\delta$ be some threshold in meters. Then we define the common message set as \vspace{-0.2cm}
\begin{equation}
\mathcal{M}_k = \left\lbrace j \in \mathcal{K}| \hspace{1mm} d_{jk} \leq \delta \right\rbrace.
\vspace{-0.2cm}
\end{equation}
In words, the common message set of user $k$ includes all the users who are located within a given range of user $k$. This simple design's intuition is that the interference becomes usually higher when the users are closer to each other. That is, on the one hand, the users in the common message set of user $k$ can mitigate the interference originating from this user, and on the other hand, the users in the proximity of user $k$ have potentially good channel conditions to the serving BSs' cluster of the common message of user $k$. Hence, they participate in achieving higher rates of the common message of user $k$ since the rate of the common message is determined by the weakest user.

The design of sets $\mathcal{K}_n^p, \mathcal{K}_n^c$ for RS-CMD is essential to the system performance of a C-RAN with data sharing. In \cite{8732995}, such a cluster design is dynamically adapted to each channel realization. Hence, the dynamic optimization of the serving clusters in each time-slot is performed using full CSIT. A similar procedure is also done when using generalized RS in \cite{9217249} with the help of full CSIT. However, the optimization of the serving clusters to adapt to the channel gains is not possible here due to the absence of full CSIT.
Therefore, in this work, we propose a practical stream-to-BS association procedure that is solely based on statistical CSI. Stream-based clustering based on statistical CSI alleviates the signaling overhead compared to dynamic association schemes performed on a time-slot basis. That is because the statistical CSI varies on a much larger time scale than the transmission time-slot. The stream-based clustering is described next.
\subsection{Proposed Stream-Based Clustering Using Statistical CSI}
In \cite{6920005}, a user-centric clustering algorithm to find a cluster of BSs that serves each user is proposed. The algorithm is called static clustering since the clusters do not change in each time-slot but rather with the path-loss, which depends on user location and varies on a much larger time-scale than channel realizations. In our case, each common stream is possibly decoded by multiple users, i.e., it has a multicast nature. The procedure in \cite{6920005} can not be directly applied in our scenario since it is meant for streams decoded by single users, i.e., it has a unicast nature.

Next, we describe the stream-based static clustering procedure, which considers the multicast nature of common streams and forms the clusters accordingly. Let  $A_{n,\text{max}}$ be the maximum number of streams that can be served with BS $n$. This parameter guarantees that BS $n$ is not overloaded with streams if it has good channel quality to many users, e.g., when those close to BS $n$. Note that $A_{n,\text{max}}$ depends mainly on the resources at BS $n$ such as the available maximum transmit power and the fronthaul capacity limit.

The remaining question is how to find the potential candidates of BSs for each stream to enter its serving cluster. Let $\mathcal{D}_k^p \eqdef \left\lbrace k \right\rbrace$ and $\mathcal{D}_k^c \eqdef \mathcal{M}_k$. Let $\mathcal{N}_k^p$ and $\mathcal{N}_k^c$ be the candidate clusters of BSs to serve the private and common streams of user $k$, respectively.
Since multiple users should decode each common stream, each BS $n$ in the candidate cluster $\mathcal{N}_k^c$ needs to have good channel quality to all users, which decode this particular stream. Since only statistical CSI is available at the CP, the channel's quality is measured based on the path-loss between the BSs and the users. We propose a criteria based on the collective channel quality to all users decoding a specific stream. Let $q_{n, \mathcal{D}_k^c}$ denotes the collective channel quality from BS $n$ and to all the users decoding the common stream of user $k$, $s_k^c$, with $q_{n, \mathcal{D}_k^c}$ given as $q_{n, \mathcal{D}_k^c} = \frac{1}{|\mathcal{D}_k^c|}{\sum_{j \in \mathcal{D}_k^c}}q_{n, j}$. Here, $q_{n, j}$ is the channel quality between user $j$ and BS $n$
and is inversely proportional to the path-loss between them.

The candidate clusters of BSs, serving the private and common messages of user $k$ are then given as
\begin{align}
\label{n1}
 \mathcal{N}_k^p &= \Big\{ n| \quad \underset{{m \in \mathcal{N}}}{\max} (q_{m,\mathcal{D}_k^p}) - q_{n,\mathcal{D}_k^p} \leq \mu \Big\}\\
 \label{n2}
  \mathcal{N}_k^c &= \Big\{ n| \quad \underset{{m \in \mathcal{N}}}{\max} (q_{m,\mathcal{D}_k^c}) - q_{n,\mathcal{D}_k^c} \leq \mu \Big\}.
\end{align}
The procedure starts by pooling the BSs candidates for each stream $s_k^p$, $s_k^c$ and associate streams  $s_k^p$ and $s_k^c$ with the best candidate in the sets $\mathcal{N}_k^p$ and $ \mathcal{N}_k^c$. After the first pooling round, if a BS $n$ is overloaded, the weakest streams with the worst channel qualities are excluded such that the number of streams does not exceed the allowed limit $A_{n,\text{max}}$ for BS $n$. Then another pooling round is performed to associate the remaining streams which were not assigned in the previous round. The procedure ends when all the streams associate with the BSs in their candidate set, or the maximum streams limit at all BSs are met. The output of this procedure are the sets $\left\lbrace \mathcal{K}_n^p, \mathcal{K}_n^c| \hspace{1mm} n \in \mathcal{N}\right\rbrace $. The same procedure can be extended to be used for stream-based clustering in generalized RS case as well.
\begin{algorithm}
	\caption{Stream-Based Clustering with Load Balancing Using Stochastic CSI}
	\label{MLBSC}
	\begin{algorithmic}[1]
		\renewcommand{\algorithmicrequire}{\textbf{Initialization:}}
		\renewcommand{\algorithmicensure}{\textbf{while}}
		\REQUIRE
		\STATE Set $A_{n,\text{max}}$ as maximum number of streams BS $n$ can serve
		\STATE Set $\mathcal{N}_k^p$ and $\mathcal{N}_k^p$ as the candidate serving clusters of private and common streams, respectively of user $k$, $\forall k \in \mathcal{K}$ according to \eqref{n1} and \eqref{n2}
		\STATE $\mathcal{S} = \{s_k^p, s_k^c|\, \forall k \in \mathcal{K}\}, \, \mathcal{N} = \{1,2,...,N\}, \,$ $\mathcal{K}_n^p = \emptyset, \mathcal{K}_n^c = \emptyset$ $\, \forall n \in \mathcal{N}$		
		\ENSURE $\mathcal{S}\neq\varnothing \, \cup \, \mathcal{N}\neq\varnothing$ \textbf{do}
		\STATE {\bfseries for} $ k \in \mathcal{K}, o \in \left\lbrace p, c \right\rbrace $
		\STATE $\quad$ {\bfseries if} $\mathcal{N}_k^o\neq\varnothing$
		\STATE $\quad$ $\quad$ The strongest BS $n$ from $\mathcal{N}_k^o$ enter the serving cluster of stream $s_k^{o}$
		\STATE $\quad$ $\quad$ $\mathcal{K}_n^o = \mathcal{K}_n^o \cup \left\lbrace k\right\rbrace $, $ \mathcal{N}_k^o = \mathcal{N}_k^o \backslash \{n\} $
		\STATE $\quad$ {\bfseries else}
		\STATE $\quad$ $\quad$ $\mathcal{S} = \mathcal{S} \backslash \{s_k^{o}\}$
		\STATE $\quad$ {\bfseries end}
		\STATE {\bfseries end}
		\STATE {\bfseries for} $n\in \mathcal{N}$
		\STATE $\quad$ {\bfseries if}$\quad$ The number of associated streams with BS $n$ exceed the maximum load, i.e., $|\mathcal{K}_n^p| + |\mathcal{K}_n^c| > A_{n,\text{max}}$
		\STATE $\quad$ $\quad$ Remove the $( |\mathcal{K}_n^p| + |\mathcal{K}_n^c| - A_{n,\text{max}} )$ weakest streams.
		\STATE $\quad$ $\quad$ $\mathcal{N} = \mathcal{N} \backslash \{n\}$, \, $ \mathcal{N}_k^p = \mathcal{N}_k^p \backslash\{n\}, \, \hspace{.5mm}\text{and} \, \hspace{.5mm} \mathcal{N}_k^c = \mathcal{N}_k^c \backslash\{n\}  \quad \forall k \in \mathcal{K}$
		\STATE $\quad$ {\bfseries end}
		\STATE {\bfseries end}
		\renewcommand{\algorithmicensure}{\textbf{end while}}
		\ENSURE
	\end{algorithmic}
\end{algorithm}

\section{Problem Formulation}\label{sec3}
\subsection{Receiver Model}
Based on the discussion in section II we can write the received signal at user $k$ as described in \eqref{e0} as
\begin{equation}\label{eq:S0}
y_k =  \underbrace{\left(\mathbf{h}_{k}^{\dagger}\mathbf{w}_{k}^p {s}_{k}^p + \sum_{j \in \Phi_{k}}\mathbf{h}_{k}^{\dagger}\mathbf{w}_{j}^c {s}_{j}^c\right)}_{\text{Signals to be decoded}} + \underbrace{\sum_{j \in \mathcal{K}\setminus k}\mathbf{h}_{k}^{\dagger}\mathbf{w}_{j}^p {s}_{j}^p + \sum_{l \in \Omega_k}\mathbf{h}_{k}^{\dagger}\mathbf{w}_{l}^c {s}_{l}^c+ n_k.}_\text{Interference plus noise}
\end{equation}
Here, $\mathbf{w}_{k}^p \in \mathbb{C}^{NL \times 1} = \left[(\mathbf{w}_{1,k}^p)^T,\ldots, (\mathbf{w}_{N,k}^p)^T \right]^T,  \mathbf{w}_{k}^c = \left[(\mathbf{w}_{1,k}^c)^T,\ldots, (\mathbf{w}_{N,k}^c)^T \right]^T$ are the aggregate beamforming vectors from all BSs to serve streams $s^{p}$ and $s^{c}$, respectively. The set $\Phi_{k}$ represents the set of common messages that user $k$ decodes and it is defined as \vspace{-0.2cm}
\begin{align}
\Phi_{k} \defeq \left\lbrace j \in \mathcal{K}|\hspace{1mm} k \in \mathcal{M}_{j} \right\rbrace. \\[-1.1cm] \nonumber
\end{align}
The set $\Omega_k$ contains the indices of users whose common messages are not decoded at user $k$.
\subsection{Achievable Rates}
The users can estimate their channels perfectly during the downlink training phase. Based on this knowledge, each user employs a SIC strategy to decode the intended messages. We assume the decoding order as follows: each user decodes the common messages first and the private message at last. In this way, part of the interference on the private message is removed. However,  to decode the common messages in the set $\Phi_k$, user $k$ follows a decoding order which is defined as\vspace{-0.2cm}
	\begin{equation} \label{eq:e10}
\pi_{k}(j):  \Phi_{k} \rightarrow \left\lbrace 1, 2, \ldots,\left|\Phi_{k}\right|  \right\rbrace, \nonumber
\vspace{-0.2cm}
\end{equation}
which represents a bijective function of the set $\Phi_{k}$ with cardinality $\left|\Phi_{k}\right|$, i.e., $\pi_{k}(j)$ is the successive decoding step in which the message $j \in \Phi_k$ is decoded at user $k$. In other terms, $\pi_{k}(j_1) > \pi_{k}(j_2)$ (where $j_1 \neq j_2$) implies that user $k$ decodes the common message of user $j_1$ first, and then the common message of user $j_2$.
Determining the optimal decoding order for each user is a complicated task and can be computationally prohibitive.

This paper proposes to fix the decoding order as follows: user $k$ starts to decode the streams based on the distance. Hence, the common messages of users closer to user $k$ are decoded before the common messages of users that are more distant from user $k$. That is, the common message of user $i$ is decoded before the common message of user $j$ if $d_{ik} < d_{jk}$.

Let $\gamma_{k}^p, \gamma_{k,i}^c$ denote the signal to interference plus noise ratios (SINR's) of user $k$, when decoding its private message and the common message of user $i$, respectively, where $\mathbf{w} =  \left[(\mathbf{w}_{1}^p)^T,  (\mathbf{w}_{1}^c)^T,\ldots,\mathbf{w}_{K}^p)^T,  (\mathbf{w}_{K}^c)^T \right]^T$. Based on equation \eqref{eq:S0}, we can write \vspace{-0.1cm}
\begin{align}
\label{eq:S1}
\gamma_{k}^p &= \frac{\left|\mathbf{h}_{k}^{\dagger}\mathbf{w}_{k}^p \right|^2}{\sigma^2 + \sum\limits_{j \in \mathcal{K}\setminus k}\left|\mathbf{h}_{k}^{\dagger}\mathbf{w}_{j}^p \right|^2 + \sum\limits_{l \in \Omega_k}\left|\mathbf{h}_{k}^{\dagger}\mathbf{w}_{l}^c \right|^2}\\
\label{eq:S2}
\gamma_{i, k}^c &= \frac{\left|\mathbf{h}_{k}^{\dagger}\mathbf{w}_{i}^c \right|^2}{\sigma^2+\sum_{j \in \mathcal{K}}\left|\mathbf{h}_{k}^{\dagger}\mathbf{w}_{j}^p \right|^2 + \sum\limits_{l \in \Omega_k}\left|\mathbf{h}_{k}^{\dagger}\mathbf{w}_{l}^c \right|^2 + \sum\limits_{m  \in \Psi_{i, k}}\left|\mathbf{h}_{k}^{\dagger}\mathbf{w}_{m}^c \right|^2} \, \raisepunct{,} 
\end{align}
where $\Psi_{i, k} \triangleq \left\lbrace m \in \Phi_k |\quad \pi_{k}(m)> \pi_{k}(i) \right\rbrace$.

The instantaneous achievable rate of user $k$ is given as $R_k = R_{k}^p + R_{k}^c$, where the instantaneous private and common rates satisfy the following achievability conditions
\vspace{-0.1cm}
\begin{align}
\gamma_{k}^p &\geq 2^{R_{k}^p/B} - 1, \quad \forall k \in \mathcal{K}, \label{eq:S3}\\
\gamma_{k,i}^c &\geq 2^{R_{k}^c/B} - 1, \quad \forall i \in \mathcal{M}_k \hspace{1.5mm} \text{and} \hspace{1.5mm} \forall k \in \mathcal{K}. \label{eq:S4}\\[-1.1cm]\nonumber
\end{align}
Note that the achievability constraint in \eqref{eq:S4} is in the form of multicasting transmission. This is the main characteristic of RS-CMD. Hence, all users in the set $\mathcal{M}_k$ which decode the common message of user $k$, i.e., $s_k^c$ must meet condition \eqref{eq:S4}. Note that the interference due to sending the common message $s_k^c$ does not affect the users in $\mathcal{M}_k$ as they also decode this message. This is the main motivation for RS-CMD in networks that suffer from interference.
\subsection{WMMSE-Rate Relationship}
We can define the MSE for user $k$ when decoding the common message of user $i$ after removing all the common messages of other users which have a decoding order less than that of user $i$ as $e_{i,k}^c \triangleq \mathbb{E} \Big\{  u_{i, k}^c (y_k-\sum\limits_{m  \in \Psi_{k}}\mathbf{h}_{k}^{\dagger}\mathbf{w}_{m}^c s_m^c) - s_i^c \Big\} $. Where $u_{i, k}^c$ is the linear receiver coefficient used at user $k$ to decode the common message of user $i$. After removing all the common streams that user $k$ decodes, we can write the MSE of the private message as $e_{k}^p = \mathbb{E}\left\lbrace u_{k}^p (y_k -\sum_{j \in \Phi_{k}}\mathbf{h}_{k}^{\dagger}\mathbf{w}_{j}^c {s}_{j}^c) - s_k^p \right\rbrace$. By using equation \eqref{eq:S0} we can write the MSE's as
\begin{align}
e_k^p & = \big|u_k^p \big|^2 T_k^p - 2\Re{\left\lbrace u_k^p \mathbf{h}_{k}^{\dagger}\mathbf{w}_{k}^p \right\rbrace} + 1 \label{me1}\\
e_{i,k}^c & = \big|u_{i, k}^c \big|^2 T_{i, k}^c - 2\Re{\left\lbrace u_{i, k}^c \mathbf{h}_{k}^{\dagger}\mathbf{w}_{i}^c \right\rbrace} + 1. \label{me2}\nonumber
\intertext{Here $T_k^p$ and $T_{i, k}^c$ are defined as}
T_k^p & = \left|\mathbf{h}_{k}^{\dagger}\mathbf{w}_{k}^p \right|^2 +  \underbrace{\sum\limits_{j \in \mathcal{K}\setminus k}\left|\mathbf{h}_{k}^{\dagger}\mathbf{w}_{j}^p \right|^2 + \sum\limits_{l \in \Omega_k}\left|\mathbf{h}_{k}^{\dagger}\mathbf{w}_{l}^c \right|^2  + \sigma^2}_{I_k^p}
\end{align}
\begin{align}
T_{i, k}^c & = \left|\mathbf{h}_{k}^{\dagger}\mathbf{w}_{i}^c \right|^2+\underbrace{\sum_{j \in \mathcal{K}}\left|\mathbf{h}_{k}^{\dagger}\mathbf{w}_{j}^p \right|^2 + \sum\limits_{l \in \Omega_k}\left|\mathbf{h}_{k}^{\dagger}\mathbf{w}_{l}^c \right|^2 + \sum\limits_{m  \in \Psi_{i, k}}\left|\mathbf{h}_{k}^{\dagger}\mathbf{w}_{m}^c \right|^2 +\sigma^2}_{I_{i, k}^c},
\end{align}
where $I_k^p$ and $I_{i, k}^c$ are the interference-plus-noise at user $k$ when decoding its private message and the common message of user $i$, respectively.
The optimal receiver coefficients which minimize the MSE's at user $k$ when decoding the private message and the common message of user $i$, respectively, are given as
\vspace{-0.5cm}
\begin{align}
u_{k, \text{mmse}}^p & = \frac{(\mathbf{w}_{k}^p)^{\dagger}\mathbf{h}_{k}}{T_k^p} \label{mmse1} \\
u_{i, k, \text{mmse}}^c & = \frac{(\mathbf{w}_{i}^c)^{\dagger}\mathbf{h}_{k}}{T_{i, k}^c} \cdot\label{mmse2}\\[-1.1cm]\nonumber
\end{align}
The expressions for MMSE receiver coefficients in \eqref{mmse1} and \eqref{mmse2} are obtained by solving the equations $\frac{\partial e_k^p}{\partial u_{k}^p} = 0$ and $\frac{\partial e_{i,k}^c}{\partial u_{i, k}^c} = 0$, respectively. By substituting the expressions of MMSE receivers in \eqref{me1} and \eqref{me2} we get the following MMSE error expressions \vspace{-0.1cm}
\begin{align}
e_{k, \text{mmse}}^p & = \frac{I_k^p}{T_k^p} \label{mse1} \\
e_{i, k, \text{mmse}}^c & = \frac{I_{i, k}^c}{T_{i, k}^c} \cdot\label{mse2}\\[-1.1cm]\nonumber
\end{align}
The critical observation here is the equivalence relation between the achievable rates and the WMSE's. The following proposition state this relation.
\proposition
The maximum achievable rate of user $k$ when decoding its private message and for user $i$ when decoding the common message of user $k$ can be expressed as
\begin{align}
\log_2(1+\gamma_{k}^p) &= \max_{u_{k}^p, \rho_{k}^p}\Big(\frac{\log(\rho_{k}^p) - \rho_{k}^p e_{k}^p + 1}{\log(2)}\Big)\label{rateMSE1} \\
\log_2(1+\gamma_{k,i}^c) &= \max_{u_{k,i}^c, \rho_{k,i}^c}\Big(\frac{\log(\rho_{k,i}^c) - \rho_{k,i}^c e_{k,i}^c + 1}{\log(2)}\Big)\cdot \label{rateMSE2}
\end{align}
To prove proposition 1, we use first-order optimality conditions to the right-hand side, unconstrained optimization problems in \eqref{rateMSE1} and \eqref{rateMSE2}, respectively. We show the equivalence in \eqref{rateMSE1}, and the same procedure can be done for \eqref{rateMSE2}. By taking the partial derivative of the objective in \eqref{rateMSE1} with respect to $u_{k}^p$ and setting the result to zero, we find out that the optimal receiver is exactly the MMSE receiver as defined in \eqref{mmse1}, i.e., $(u_{k}^{p})^* = u_{k, \text{mmse}}^p$. By taking the derivative with respect to $\rho_{k}^p$ and set it to zero, we get the optimal coefficient $(\rho_{k}^{p})^* = \frac{1}{e_{k, \text{mmse}}^p}$. By substituting $u_{k}^{p}$ and $\rho_{k}^{p}$ in the right-hand side expression of \eqref{rateMSE1} with $u_{k, \text{mmse}}^p$ and $\frac{1}{e_{k, \text{mmse}}^p}$, respectively, we get exactly the maximum achievable rate expression on the left-hand side of \eqref{rateMSE1}. The same procedure can be followed to show the equivalence in \eqref{rateMSE2}.
The rate-MSE relation in proposition 1 proves to be very helpful when discussing the WMMSE-based optimization algorithm later in section \ref{sec4}.
Note that the SINR and rate expressions in equations \eqref{eq:S3}, \eqref{eq:S4}, respectively, are deterministic functions when the CP has perfect CSI. However, when imperfect CSI is assumed, and the vector $\mathbf{h}$ is random, the SINR and the rate are random functions. Next, we discuss how to characterize the achievable rates and the corresponding MSE in the imperfect CSI scenario.
\subsection{Ergodic Rate and WMSE}
For achieving the instantaneous rates (and the corresponding WMSE's) as given in equations \eqref{eq:S3} and \eqref{eq:S4}, we need full CSIT knowledge at the CP. Moreover, the rate is a function of the beamforming vector, which we need to adapt to the channel realization in each coherence block. The availability of full CSIT is rather an optimistic assumption. In a practical scenario, the uncertainty in CSIT is inevitable. With full CSIT, we can perform system level optimization to maximize the deterministic function $R_k = R_{k}^p + R_{k}^c$. However, CSIT uncertainty brings technical difficulties because of the expressions in \eqref{eq:S3} and \eqref{eq:S4} can not be directly used due to the lack of perfect knowledge of the channel coefficients. Hence, the rate expressions in \eqref{eq:S3} and \eqref{eq:S4} are random and depends on the specific channel realization, which is not available at the transmitter. 

To overcome the difficulty of optimizing the system performance under CSIT uncertainty, we adopt in this paper a stochastic optimization approach that seeks to optimize the performance on average. To this end, we define the achievable ergodic rate of user $k$ in this case as $\mathbb{E}_{\mathbf{h}}\left\lbrace R_{k}^{p} + R_{k}^{c} \right\rbrace \triangleq \bar{R}_{k}^{p} + \bar{R}_{k}^{c}$, where the expectation operator is with respect to the random channel and $\bar{R}_{k}^{p}$, $\bar{R}_{k}^{c}$ are the ergodic private and common rates, respectively of user $k$. By taking the expectation of both sides of equations \eqref{rateMSE1} and \eqref{rateMSE2} we get the following relation between the ergodic rate and the ergodic WMSE's
 \vspace{-0.4cm}
\begin{align}
\mathbb{E}_{\mathbf{h}}\left\lbrace\log_2(1+\gamma_{k}^p)\right\rbrace &= \frac{1}{\log(2)}\mathbb{E}_{\mathbf{h}}\{\max_{u_{k}^p, \rho_{k}^p}\big(\log(\rho_{k}^p) - \rho_{k}^p e_{k}^p + 1\big)\}\label{ErateMSE1} \\
\mathbb{E}_{\mathbf{h}}\left\lbrace\log_2(1+\gamma_{k,i}^c)\right\rbrace &= \frac{1}{\log(2)}\mathbb{E}_{\mathbf{h}}\{\max_{u_{k,i}^c, \rho_{k,i}^c}\big(\log(\rho_{k,i}^c) - \rho_{k,i}^c e_{k,i}^c + 1\big)\}\label{ErateMSE2} \\[-1.1cm]\nonumber
\end{align}
The expectation operator on the right-hand side of \eqref{ErateMSE1} and \eqref{ErateMSE2} is outside the maximum operator to make it explicit the dependency of the MSE weight and receiver coefficients on the channel realizations. Next, we formulate the optimization problem which we want to investigate in this paper.
\subsection{ESR Maximization}
In this work, we are interested in the joint optimization of stochastic coordinated beamforming (SCB) and rate allocation for each user's private and common streams so that the ergodic sum-rate is maximized. The maximization is done subject to per-BS fronthaul and power constraints and ergodic rates achievability constraints.
The problem can then be mathematically expressed as
\vspace{-0.5cm}
\begin{subequations}\label{eq:Opt0}
	\begin{align}
	\mathcal{P}_{0}:\quad&\underset{\mathbf{w}, \bar{\mathbf{R}}}{\text{maximize}}\quad \sum_{k \in  \mathcal{K}}\left( \bar{R}_{k}^{p} + \bar{R}_{k}^{c} \right)         \label{eq:Obj0} \\
	&\text{subject to} \quad \nonumber\\ 
	&\sum_{{k \in \mathcal{K}_n^p}}\normV[\big]{\mathbf{w}_{n ,k}^{p}}_2^2  + \sum_{{k \in \mathcal{K}_n^c}}\normV[\big]{\mathbf{w}_{n ,k}^{c}}_2^2\leq P_{n}^{\text{Max}} \quad \forall n \in \mathcal{N} \label{Power}\\
	& \sum_{{k \in \mathcal{K}_n^p}}\bar{R}_{k}^{p} + \sum_{{k \in \mathcal{K}_n^c}} \bar{R}_{k}^{c} \leq F_n \quad \forall n \in \mathcal{N} \label{Fronthaul}\\
	& \bar{R}_{k}^{p} \leq \mathbb{E}_{\mathbf{h}}\left\lbrace B\log_2(1+\gamma_{k}^p)\right\rbrace \quad \forall k \in \mathcal{K} \label{priv} \\
		& \bar{R}_{k}^{c} \leq \mathbb{E}_{\mathbf{h}}\left\lbrace B\log_2(1+\gamma_{k, i}^c)\right\rbrace  \quad \forall i \in \mathcal{M}_k \hspace{1.5mm} \text{and} \hspace{1.5mm} \forall k \in \mathcal{K}. \label{comm}
	\end{align}
\end{subequations}
Here, $\bar{\mathbf{R}}$ is a set of variables $\bar{\mathbf{R}} \triangleq\left\lbrace \bar{R}_{k}^p, \bar{R}_{k}^c|\quad \forall k \in \mathcal{K}\right\rbrace$ which represents the ergodic private and common rates of all users. With SCB optimization, the same beamforming vectors $\mathbf{w}$ are used for all transmit blocks in which the distribution of the channel remains constant. Problem $\mathcal{P}_0 $ is difficult and challenging to solve. In particular, the constraints \eqref{priv} and \eqref{comm} are functions of a stochastic quantity.
Hence, the achievable rates depend on the current realization of channel fading, which is unknown at the CP, and the expected value in \eqref{priv} and \eqref{comm} has no closed-form. Moreover, even when considering the deterministic version of this problem (e.g., assuming perfect CSIT as in \cite{8732995}), the SINR is a non-convex function of the design variables (i.e., the beamforming vectors), and the resulting problem is known to be NP-hard.
\section{Optimization Algorithm}\label{sec4}
 To tackle this problem, we propose a two-step approach. First, we approximate the ergodic rate expression using the equivalence between the expected value of a stochastic process and its time-average, where we use the sample average approximation (SAA) method. After that, we use the WMSE-rate relationship to develop a WMMSE-based algorithm. The algorithm has polynomial complexity, and it is guaranteed to obtain a stationary solution to the ESR problem. Next, we discuss the SAA reformulation of the problem $\mathcal{P}_0$.

%
%
%
%
\subsection{SAA Approach}
To approximate the stochastic function in \eqref{priv} and \eqref{comm}, we propose to use the SAA approach \cite{StochSh}. Let $M \in \mathbb{N}$ denotes the sample size. We define the SAA of the achievable ergodic rate as \vspace{-0.1cm}
\begin{equation}\label{SAA}
\bar{R}_{k} \leq  \frac{B}{M}\sum_{m = 1}^{M} \left( \log_2\left( 1 + \gamma_{k}^p(m)\right) +  \min_{i \in \mathcal{M}_k}\log_2\left( 1 + \gamma_{k, i}^c(m)\right) \right).
\vspace{-0.2cm}
\end{equation}
Not that in \eqref{SAA} we write $\gamma_{k}^p(m), \gamma_{k, i}^c(m)$ to emphasis on the dependency of SINR on the channel sample $\mathbf{h}^m$. Here, $\left\lbrace \mathbf{h}^m|\hspace{1mm} 1\leq m \leq M\right\rbrace$ are independent realizations of the random vector $\mathbf{h}$ which represents the aggregate channels vector of all users and is defined as $\mathbf{h} = \left[   \mathbf{h}_{1}^T,\ldots, \mathbf{h}_{K}^T \right]^T\in \mathbb{C}^{NLK\times 1}$. Now, we reformulate problem $\mathcal{P}_0$ using the SAA as \vspace{-0.4cm}
\begin{subequations}\label{eq:Opt1}
	\begin{align}
	\mathcal{P}_{1}(M):\quad&\underset{\mathbf{w}, \mathbf{R}}{\text{maximize}}\quad  \sum_{k \in  \mathcal{K}} \left(\bar{R}_{k}^p  + \bar{R}_{k}^c\right)  \label{eq:Obj4} \\
	&\text{subject to} \quad  \eqref{Power}, \eqref{Fronthaul} \nonumber \\ 
& \bar{R}_{k}^p -  \frac{B}{M}\sum_{m = 1}^{M} \log_2\left( 1 + \gamma_{k}^p(m)\right) \leq 0 \quad \forall k \in \mathcal{K} \label{priv1}\\
& \bar{R}_{k}^c - \frac{B}{M}\sum_{m = 1}^{M} \log_2\left( 1 + \gamma_{k, i}^c(m)\right) \leq 0 \quad \forall i \in \mathcal{M}_k, \forall k \in \mathcal{K}. \label{comm1}
	\end{align}
\end{subequations}
Obviously, problem $\mathcal{P}_1(M)$ depends on the sample size $M$. Although
problem $\mathcal{P}_1(M)$ is non-convex, it is more tractable than problem $\mathcal{P}_0$ as it is deterministic thanks to SAA. Moreover, the deterministic approximation using the SAA method does not lose any optimality of problem $\mathcal{P}_0$ in the asymptomatic regime as indicated by the following theorem
\theorem  The set of globally optimal solutions of problem $\mathcal{P}_1(M)$ converges to the set of globally optimal solutions of the stochastic problem $\mathcal{P}_0$ in the asymptotic regime when $M \rightarrow \infty$.
\begin{proof}
	Please refer to the appendix $A$ for further details.
\end{proof}
\vspace{-0.5cm}
 \subsection{WMMSE Based Algorithm}
Problem $\mathcal{P}_1(M)$ is non-convex due to constraints in \eqref{priv1} and \eqref{comm1}. Moreover, it depends on the Monte-Carlo sample size $M$, which increases the computation complexity. In \cite{7555358}, the authors use a conventional WMMSE algorithm to solve the ESR maximization problem for a specific RS scheme in MISO-BC. In the conventional WMMSE algorithm \cite{6623158}, the WMSE-rate relation is used to formulate an equivalent MSE minimization problem that replaces the achievable rate variables with MSE variables. However, this approach can not be applied here due to the limited fronthaul capacity constraints in which the rate variables are explicitly used. To overcome this difficulty we use in this work the WMMSE-rate relationship as shown in \eqref{rateMSE1} and \eqref{rateMSE2} directly in the constraints. Let us define the set of variables
\vspace{-0.6cm}
\begin{align}
 & \bm \rho= \left\lbrace \rho_k^p(m), \rho_{k, i}^c(m)| \quad  \forall i \in \mathcal{M}_k, \forall k \in \mathcal{K}, \forall m \hspace{1mm}1\leq m \leq M\right\rbrace \\ 
 & \mathbf{u} = \left\lbrace u_k^p(m), u_{k, i}^c(m)| \quad  \forall i \in \mathcal{M}_k, \forall k \in \mathcal{K}, \forall m \hspace{1mm}1\leq m \leq M \right\rbrace. 
\end{align}
Problem $\mathcal{P}_1(M)$ can be reformulated based on the WMMSE-rate equivalence as follows \vspace{-0.2cm}
\begin{subequations}\label{eq:Opt2}
	\begin{align}
	\mathcal{P}_{2}(M):\quad&\underset{\mathbf{w}, \mathbf{u},\mathbf{R}, \bm \rho }{\text{maximize}}\quad  \sum_{k \in  \mathcal{K}} \left(\bar{R}_{k}^p  + \bar{R}_{k}^c\right)  \label{eq:Obj6} \\
	&\text{subject to} \quad  \eqref{Power}, \eqref{Fronthaul} \nonumber \\
& \bar{R}_{k}^p -  \frac{B}{M}\sum_{m = 1}^{M}\max_{u_{k}^p(m), \rho_{k}^p(m)}\Big(\frac{\log(\rho_{k}^p(m)) - \rho_{k}^p(m) e_{k}^p(m) + 1}{\log(2)}\Big) \leq 0 \label{priv4}\\
& \bar{R}_{k}^c - \frac{B}{M}\sum_{m = 1}^{M} \max_{u_{k,i}^c(m), \rho_{k,i}^c(m)}\Big(\frac{\log(\rho_{k,i}^c(m)) - \rho_{k,i}^c(m) e_{k,i}^c(m) + 1}{\log(2)}\Big) \leq 0. \label{comm4}
\vspace{-0.2cm}
	\end{align}
\end{subequations}
Here, the dependency of the MSE weight and receiver coefficients on the channel realization $m$ is made explicit. The rationale behind using WMMSE is twofold. First, we note that for a fixed set of variables $\mathbf{u},\bm \rho$ the optimization problem is convex with respect to the variables $\mathbf{w},\mathbf{R}$. Moreover, when the variables  $\mathbf{w},\mathbf{R}$ are fixed the optimal values of $\mathbf{u},\bm \rho$ can be found in closed-form. Second, with the WMMSE reformulation, we can avoid optimization algorithms' dependence on the sample size $M$ as we see next. For fixed $\mathbf{w},\mathbf{R}$, we can compute the optimal receiver and MSE weight coefficients using \eqref{mmse1}, \eqref{mmse2}, \eqref{mse1} and \eqref{mse2}. Let us first define the following set of coefficients, which we are going to use in the WMMSE-based algorithm
\begin{align}
&u_{k}^p(m) = \frac{(\mathbf{w}_{k}^p)^{\dagger}\mathbf{h}_{k}^m}{T_k^p(m)}, \quad  u_{k, i}^c(m) = \frac{(\mathbf{w}_{k}^c)^{\dagger}\mathbf{h}_{i}^m}{T_{k, i}^c(m)} \label{e2},\\
&\rho_{k}^p(m) = 1/e_{k, \text{mmse}}^p(m), \quad \rho_{k, i}^c(m) = 1/e_{k, i, \text{mmse}}^c(m) \label{cc1},\\
&\bar{t}_{k}^p = \frac{1}{M} \sum_{m = 1}^M \rho_{k}^p(m)\normV[\big]{u_{k}^p(m)}_2^2, \quad \bar{t}_{k,i}^c = \frac{1}{M} \sum_{m = 1}^M \rho_{k, i}^c(m)\normV[\big]{u_{k, i}^c(m)}_2^2, \label{cc2}\\
&\bar{z}_{k}^p = \frac{1}{M} \sum_{m = 1}^M(1 - \rho_{k}^p(m) + \log(\rho_{k}^p(m))) , \quad \bar{z}_{k, i}^c = \frac{1}{M} \sum_{m = 1}^M(1 - \rho_{k, i}^c(m) + \log(\rho_{k, i}^c(m))),\\
&\bar{\mathbf{f}}_{k}^p = \frac{1}{M} \sum_{m = 1}^M \rho_{k}^p(m)\mathbf{h}_k^m (u_{k}^p(m))^{\dagger}, \quad \bar{\mathbf{f}}_{k,i}^c = \frac{1}{M} \sum_{m = 1}^M \rho_{k, i}^c(m)\mathbf{h}_i^m (u_{k, i}^c(m))^{\dagger},\\
&\bar{\mathbf{Y}}_{k,k}^p = \frac{1}{M} \sum_{m = 1}^M\big(\rho_{k}^p(m)\normV[\big]{u_{k}^p(m)}_2^2 \mathbf{h}_k^m (\mathbf{h}_k^m)^{\dagger} \big),\\
&\bar{\mathbf{Y}}_{k,i}^c = \frac{1}{M} \sum_{m = 1}^M\big(\rho_{k, i}^c(m)\normV[\big]{u_{k, i}^c(m)}_2^2 \mathbf{h}_i^m (\mathbf{h}_i^m)^{\dagger} \big). \hspace{6.5cm} \label{e8}
\end{align}
By plugging the optimal values into constraints \eqref{priv4} and \eqref{comm4} and using \eqref{me1} and \eqref{me2}, we get the following optimization problem over $\mathbf{w},\mathbf{R}$.
\begin{subequations}\label{eq:Opt3}
	\begin{align}
	\mathcal{P}_{3}:\quad&\underset{\mathbf{w}, \mathbf{R} }{\text{maximize}}\quad  \sum_{k \in  \mathcal{K}} \left(\bar{R}_{k}^p  + \bar{R}_{k}^c\right)  \label{eq:Obj7} \\
	&\text{subject to} \quad  \eqref{Power}, \eqref{Fronthaul} \nonumber \\
& \sum_{j \in \mathcal{K}}(\mathbf{w}_j^p)^{\dagger}\bar{\mathbf{Y}}_{k,k}^p \mathbf{w}_j^p + \sum_{l \in \Omega_k}(\mathbf{w}_l^c)^{\dagger}\bar{\mathbf{Y}}_{k,k}^p \mathbf{w}_l^c  - 2\Re\left\lbrace (\bar{\mathbf{f}}_{k}^p)^{\dagger}\mathbf{w}_k^p \right\rbrace + \nonumber \\& \frac{\log(2)\bar{R}_{k}^p}{B} + \sigma^2\bar{t}_{k}^p  - \bar{z}_{k}^p \leq 0 \quad \forall k \in \mathcal{K} \label{priv7}\\
& \sum_{j \in \mathcal{K}}(\mathbf{w}_j^p)^{\dagger}\bar{\mathbf{Y}}_{k,i}^c \mathbf{w}_j^p + \sum_{l \in \Omega_i}(\mathbf{w}_l^c)^{\dagger}\bar{\mathbf{Y}}_{k,i}^c \mathbf{w}_l^c + \sum_{m \in \Psi_{k, i}}(\mathbf{w}_m^c)^{\dagger}\bar{\mathbf{Y}}_{k,i}^c \mathbf{w}_m^c + (\mathbf{w}_k^c)^{\dagger}\bar{\mathbf{Y}}_{k,i}^c \mathbf{w}_k^c - \nonumber \\& 2\Re\left\lbrace (\bar{\mathbf{f}}_{k, i}^c)^{\dagger}\mathbf{w}_k^c \right\rbrace +  \frac{\log(2)\bar{R}_{k}^c}{B} + \sigma^2\bar{t}_{k, i}^c  - \bar{z}_{k, i}^c \leq 0 \quad \forall i \in \mathcal{M}_k, \forall k \in \mathcal{K}. \label{comm7}
	\end{align}
\end{subequations}
Problem $\mathcal{P}_3$ is a convex optimization problem which can be efficiently solved using the interior point method, implemented in optimization solvers such as \cite{cvx}. Moreover, problem $\mathcal{P}_3$ is independent of the Monte-Carlo sample size, which reduces the computational complexity significantly compared to other optimization algorithms, as used in \cite{7185432,6612065}.
Algorithm \ref{e1} lists the steps to obtain a stationary solution of problem $\mathcal{P}_2(M)$.
\begin{algorithm}
	\caption{WMMSE-based algorithm}
	\label{e1}
	\begin{algorithmic}[1]
		\renewcommand{\algorithmicrequire}{\textbf{Step 0:}}
		\renewcommand{\algorithmicensure}{\textbf{Repeat}}
		\REQUIRE Initialize the beamforming vectors $\mathbf{w}$ to be feasible. Using the statistical CSI, generate the $M$ samples of channel vector as $\left\lbrace \mathbf{h}^1,\ldots, \mathbf{h}^M\right\rbrace $
		\ENSURE
		\STATEx \textbf{Step 1:} Update the set of auxiliary variables $\left\lbrace \bar{t}_{k}^p, \bar{t}_{k, i}^c, \bar{z}_{k}^p, \bar{z}_{k, i}^c, \bar{\mathbf{f}}_{k}^p,\bar{\mathbf{f}}_{k, i}^c, \bar{\mathbf{Y}}_{k,k}^p,\bar{\mathbf{Y}}_{k,i}^c\right\rbrace$ using equations \eqref{e2}-\eqref{e8}.
		\STATEx \textbf{Step 2:} Update the beamforming vectors and the allocated private and common average rates per-user, by solving the convex optimization problem $\mathcal{P}_3$.
		\renewcommand{\algorithmicensure}{\textbf{Until} convergance}
		\ENSURE
	\end{algorithmic}
\end{algorithm}
\vspace{-0.2cm}
\theorem \label{t2} Let $ \left\lbrace \mathbf{q}^r\right\rbrace_{r=1}^{\infty} \triangleq \left\lbrace \mathbf{w}^r, \mathbf{u}^r,\mathbf{R}^r, \bm \rho^r   \right\rbrace_{r=1}^{\infty} $ be the sequence generated by algorithm \ref{e1}, where $r$ is the iteration number. The sequence $ \left\lbrace \mathbf{q}^r\right\rbrace_{r=1}^{\infty}$ converges to a KKT solution of problem $\mathcal{P}_2(M)$.
\begin{proof}
The details are given in appendix $B$.
\end{proof}

\subsection{Complexity Analysis of Algorithm 2}
The solution provided by algorithm 2 converges to a KKT solution of the problem $\mathcal{P}_2(M)$ according to theorem \ref{t2}. The major computational load in Algorithm 2 lies in step 2. Hence, to update the beamforming vectors and allocated rate variables in step 2, we need to solve $\mathcal{P}_3$, which is a quadratically constrained convex optimization problem (QCCP). The problem $\mathcal{P}_3$ consists of $d1 = 2(N + K+\sum_{{k \in \mathcal{K}}}|\mathcal{M}_k|)$ constraints, where $|\mathcal{M}_k|$ is the cardinality of set $\mathcal{M}_k$ and $d2 = (2K(NL+1))$ variables. Hence, the worst-case computational complexity of solving the QCCP is $\mathcal{O}((d_1 d_2^2 + d_2^3)\sqrt{d_1} log(1/{\epsilon}))$ for a given solution accuracy $\epsilon$ when using an interior-point based solver \cite{doi:10.1137/1.9781611970791}. We note here that the complexity of the algorithm does not depend on the Monte-Carlo sample size $M$, which is a great advantage as compared to the difference of convex (DC) programming approach used by \cite{6996028} since $M$ can be very large for the SAA to be an accurate approximation of the stochastic program.
\section{Numerical Simulations} \label{NS}
In this section, we conduct a set of numerical simulations to evaluate the performance of the proposed RS-CMD, compared to the TIN and generalized RS (together with RS sub-schemes thereof).
For all the simulations we consider the following channel model \cite{6786060}:
\begin{equation} \label{eee3}
	h_{n, k} = D_{n, k} \mathbf{e}_{n, k}.
\end{equation}
Here, $D_{n, k} = 10^{-PL(d_{n, k})/20}\sqrt{g_{n, k} s_{n, k}}$, where $g_{n, k}$ is the shadowing coefficient, $s_{n, k}$ is the antenna gain and $PL(d_{n, k})$ is the path-loss coefficient defined as
\begin{equation}
PL(d_{n, k}) = 148.1 + 37.6\log_{10}(d_{n, k}).
\end{equation}
Here, $d_{n, k}$ is the distance between BS $n$ and user $k$ in km. The coefficients $\mathbf{e}_{n, k} \in \mathbb{C}^{L \times 1}$ in \eqref{eee3} represent the small-fading component and is modelled as $\mathbf{e}_{n, k} \sim \mathcal{CN}(0,\mathbf{I}_L)$. This work considers that the CP can only perfectly estimate the large fading coefficient $D_{n, k}$ \cite{SSUM}, however the small-fading coefficient is unknown at the CP. The noise spectral density is set to $-169$ dBm/Hz and the maximum transmit power per-BS is chosen to be $P_{n}^{\text{Max}} = 20 \hspace{1mm} \text{dBm}\, \forall n \in \mathcal{N}.$

The proposed algorithm is evaluated with different RS schemes. In addition to the generalized RS scheme we also consider TIN and NOMA as two baseline schemes. The proposed clustering procedure in algorithm \ref{MLBSC} and the optimization approach of merging SAA with the iterative algorithm \ref{e1} can be directly extended for the case of generalized RS and its sub-schemes. Due to the space limit we omit the details of specific problem formulation for the generalized RS case. Instead, we briefly discuss the system model of these schemes and the number of streams to be transmitted as a function of the number of users. In particular we consider the following schemes
\begin{itemize}
    \item[1)] \textbf{Generalized RS scheme:} In generalized RS proposed in \cite{Mao}, for a network with $K$ users, the number of common streams is given as $2^K-K-1$ which is an exponential function in the number of users. The design of common message sets does not depend on the CSIT or network topology.
	\item[2)] \textbf{RS scheme 1:} To reduce the complexity of the generalized RS scheme, we propose to use only the first and second order streams, which are either decoded by one user (private streams) or decoded at two users at most. By using this scheme we reduce the total number of streams from $2^K-1$ to $\frac{K(K+1)}{2}$.
	\vspace{1mm}
	\item[3)] \textbf{RS scheme 2:} The scheme proposed by \cite{7555358} which uses the first and last layers only from the generalized RS scheme, i.e., the number of streams in this scheme is $K+1$ streams and the common streams must be decoded by all users.
	\vspace{1mm}
	\item[4)] \textbf{NOMA:} The number of schemes in NOMA is equal to the number of users. NOMA relies on supervision coding (SC) at the transmitter and successive interference cancellation at the receivers. In the simulations we adopt the SC-SIC per group multi-antenna NOMA strategy, similar to the scheme adopted in \cite{Mao} for a MISO-BC.
	\vspace{1mm} 	
	\item[5)] \textbf{TIN scheme:} The conventional TIN scheme (the first order stream of the generalized scheme). 	
\end{itemize}
To highlight the importance of cloud computing platform and capture the practical aspects in C-RAN, we run the algorithms on a cloud computing platform. For this purpose we use the Amazon elastic cloud computing platform (Amazon EC2) provided from Amazon web services (AWS) \cite{amazon} where we have deployed the computing instance m5.2xlarge \cite{EC2}. The optimization is performed employing the MOSEK solver \cite{mosek} with CVX \cite{cvx}.
\subsection{Comparison of the Two Studied RS Strategies}
The generalized RS scheme's design does not depend on specific user's information such as locations or CSIT. Hence, the generalized RS scheme provides a robust strategy to manage the network's interference against the channel imperfections. However, it introduces a substantial computational overhead as the number of streams grows exponentially with the users' numbers. On the other hand, the number of streams used in RS-CMD increases linearly with the number of users. Specifically, the number of streams is $2K$ for the $K$-user case. The first $K$ streams contain the coded unicast (private) messages of all users, and the last $K$ streams include the coded multicast common messages of all users, which makes it scalable with the network size as opposed to the generalized RS.
RS-CMD is also robust against CSIT imperfection as it requires only the knowledge of the receiver locations, which can be accurately estimated with little communication overhead.
In this work, the generalized RS scheme is basically used as a benchmarking scheme in small size networks. For larger networks, we propose to use the practical RS-CMD scheme, which copes efficiently with the number of users. Next, we illustrate the complexity of all studied transmission schemes in terms of the number of streams.

\subsection{Complexity of the studied schemes}
\begin{figure}
   \begin{minipage}{0.49\linewidth}
	\centering
	\tikzset{every picture/.style={scale=.9}, every node/.style={scale=.9}}
	\includegraphics[width=\linewidth]{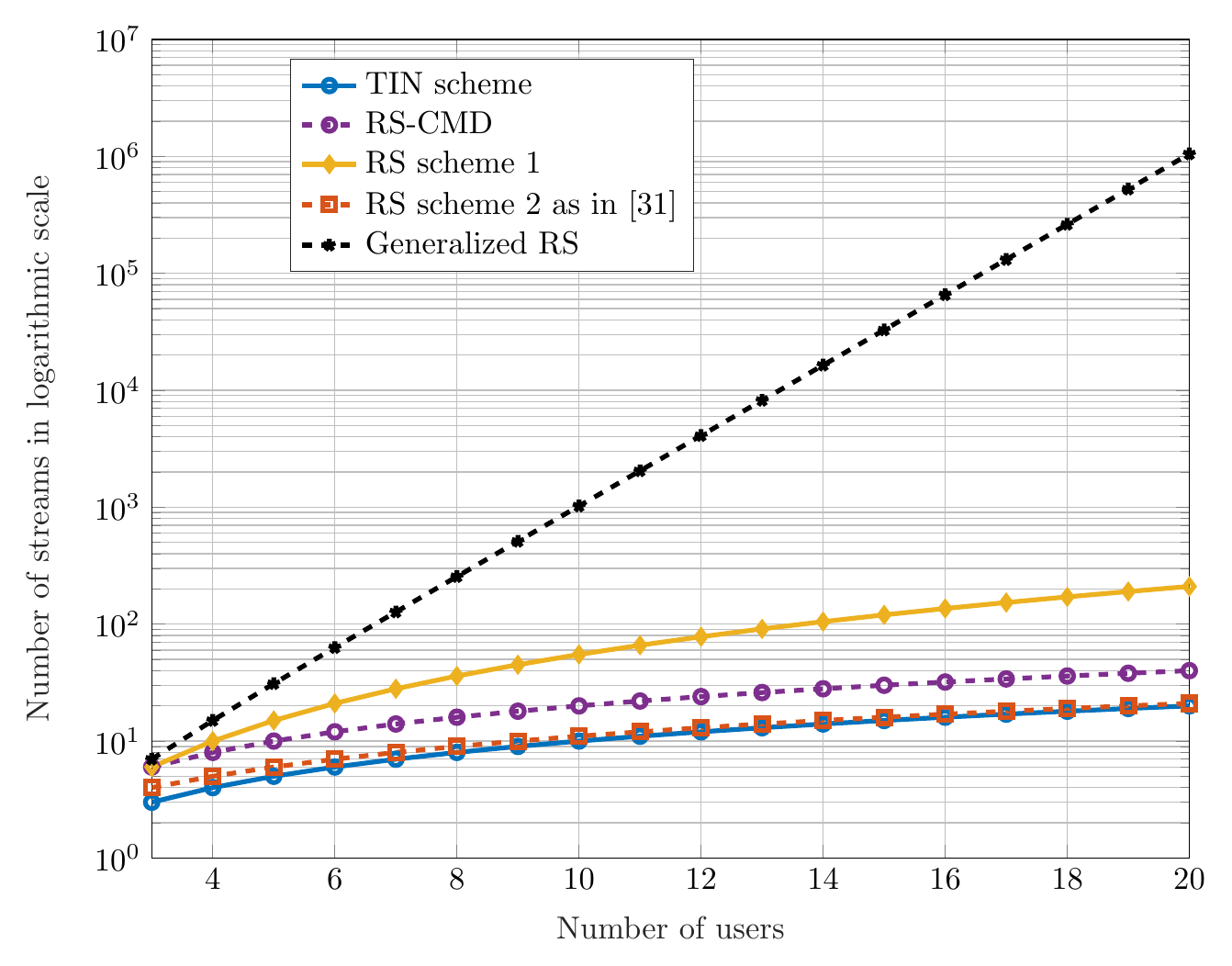}
	\caption{Complexity of the studied schemes in terms of number of streams.}
	\label{Fig1}
\end{minipage}\hfill
\begin{minipage}{0.49\linewidth}
	\centering
	\includegraphics[width=\linewidth]{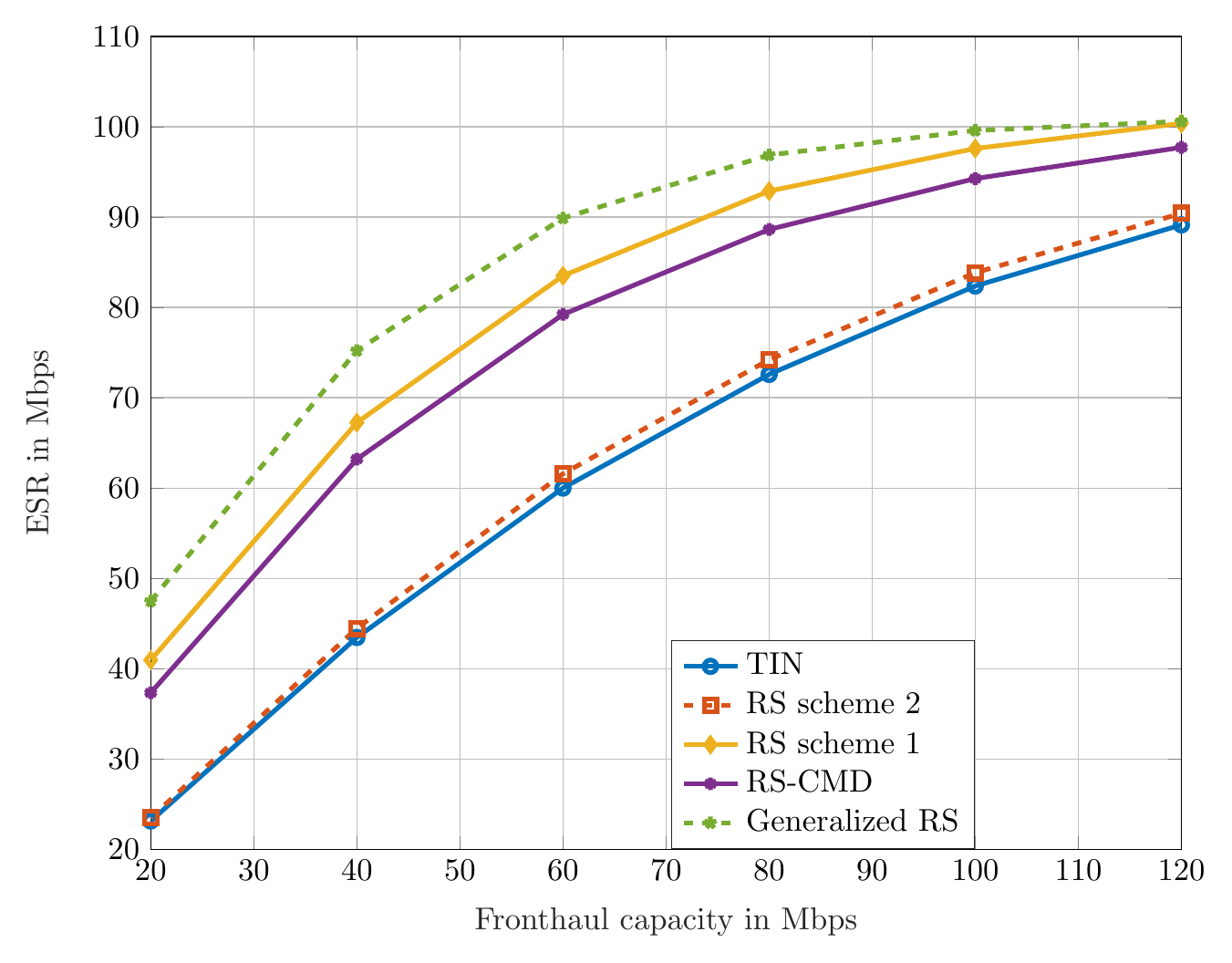}
	\caption{Ergodic sum-rate as a function of maximum fronthaul capacity per BS.}
	\label{Tex_fig}
\end{minipage}
\end{figure}

First, we illustrate the complexity of different studied RS schemes in terms of the number of streams we need to optimize. Fig.~\ref{Fig1} shows the number of streams in each scheme as a function of the number of users. {Obviously, the generalized RS scheme is computationally involved and can only be implemented for a small number of users. Hence, the number of streams in the generalized RS scheme is exponentially increasing with the number of users, which makes solving problem $\mathcal{P}_3$ computationally prohibitive for a large number of users. Even when we use the RS scheme 1, which considers the first and second-order streams, the resulting number of streams is quadratic in the number of users, making it much more complicated to optimize than RS-CMD. Next, we use the generalized RS scheme as a small network size benchmark to evaluate the proposed practical RS schemes.}
\subsection{Ergodic Sum-rate as a function of fronthaul capacity}
To explore the impact of fronthaul capacity on the achievable ergodic sum-rate, we generate a C-RAN consisting of four BSs, each with two antennas serving a set of four users. We evaluate the performance of all studied schemes using Algorithm \ref{e1} for different values of fronthaul capacities. The simulations are averaged over one-hundred network realizations. For each network realization, we generate a Monte-Carlo sample to evaluate the SAA. The sample consists of $M = 1000$ independent and identically distributed (i.i.d) channel realizations. In Fig.~\ref{Results_user_2}, the results of these simulations are depicted. In this scenario, all the RS-based schemes achieve a considerable gain in ESR terms compared to TIN. Interestingly, generalized RS, RS-CMD, and RS scheme 1 have comparable performance. However, although RS scheme 2 achieves higher ESR than TIN, it performs worse than the other studied RS schemes, especially in the regimes with low fronthaul capacity.
Interestingly, by comparing Fig.~\ref{Results_user_2} and Fig.~\ref{Fig1}, we see that the ESR performance is proportional to the number of common streams used in the transmission scheme. Hence, the extra degrees of freedom brought by common streams play an essential role in achieving a good ESR performance.
In Fig.~\ref{Fig1}, we see that when the number of users is four, the number of streams used by the schemes is comparable (apart from the generalized RS). As the number of users increases, the number of streams considerably varies between the transmission schemes. For instance, when the number of users is equal to twenty, RS scheme 1 utilizes way more streams as TIN. In contrast, the number of streams in RS-CMD is comparable with RS-scheme 2 and TIN. Hence, it is interesting to see the impact of increasing the number of users on the performance of the studied schemes. Next, we study the ESR as a function of the number of the users for all the schemes, except for the generalized RS scheme for which the computational load becomes very high as the number of users increases.
\subsection{Ergodic Sum-rate as a function of the number of users}
\begin{figure}
   \begin{minipage}{0.49\textwidth}
	\includegraphics[width=\linewidth]{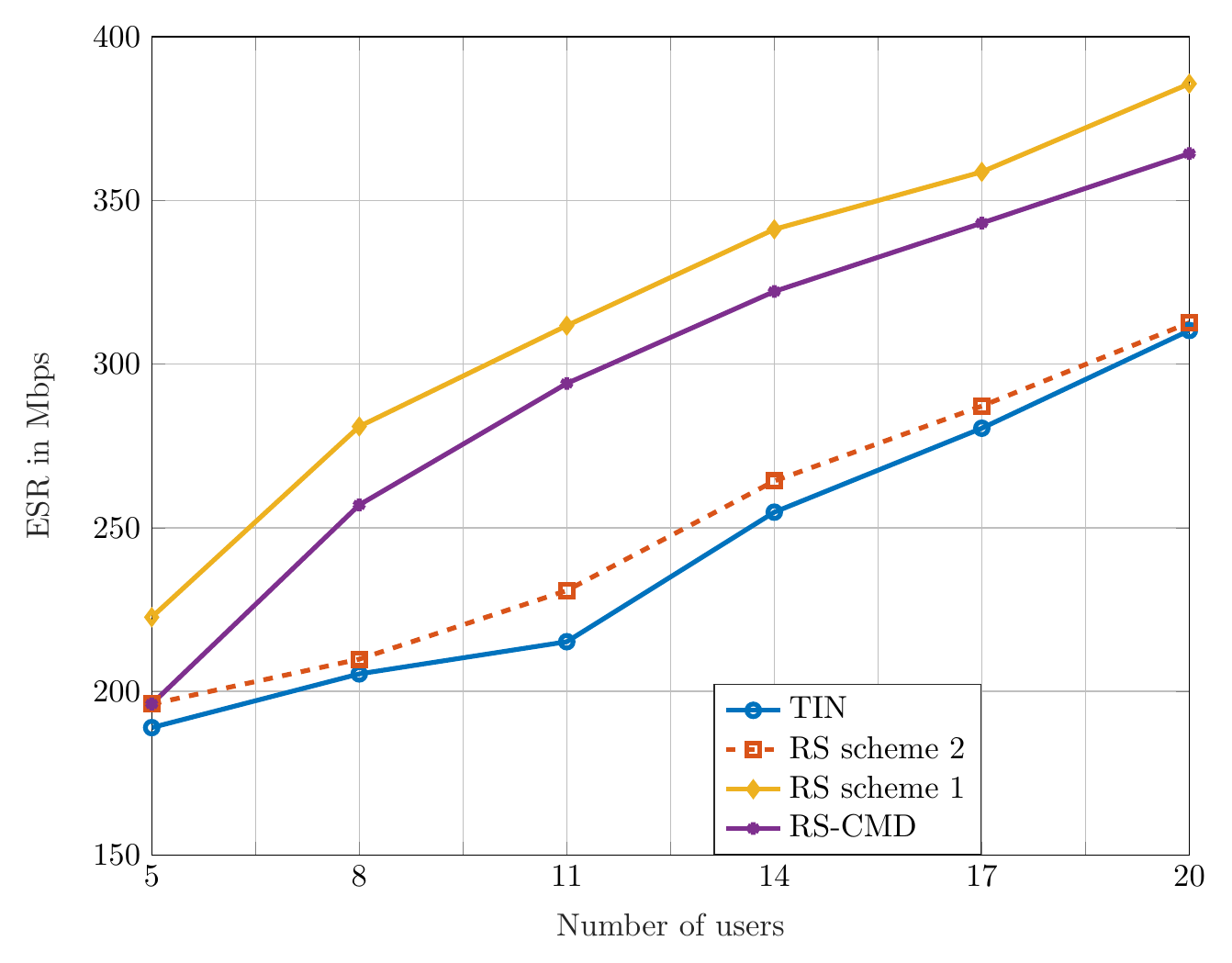}
	\caption{Ergodic sum-rate as a function of the number of users, fronthaul capacity per-BS is 200 Mbps.}
	\label{Results_user_2}
\end{minipage}\hfill
\begin{minipage}{0.49\textwidth}
	\includegraphics[width=\linewidth]{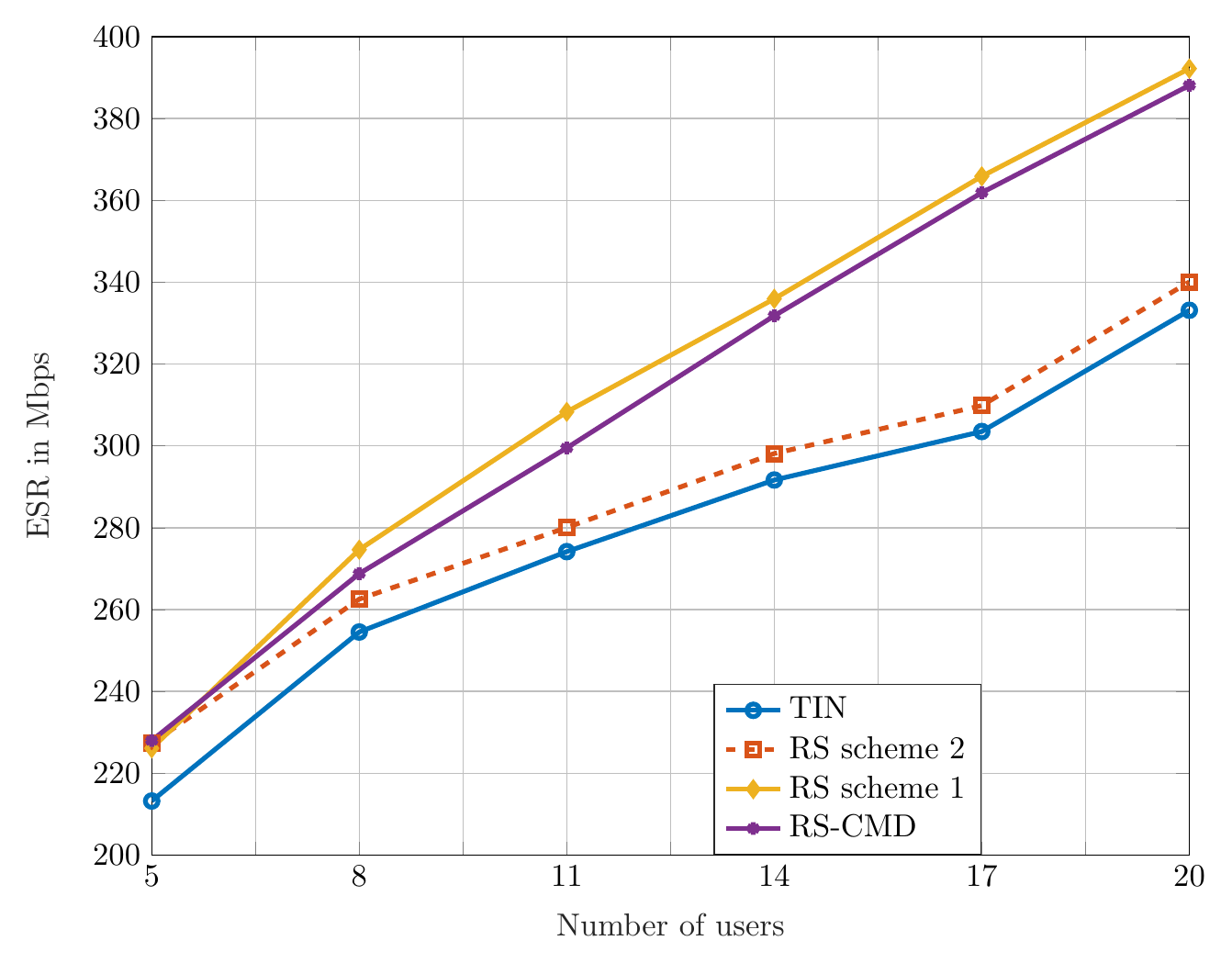}
	\caption{Ergodic sum-rate as a function of the number of users, fronthaul capacity per-BS is 300 Mbps.}
	\label{Results_user_3}
\end{minipage}
\vspace{-0.7cm}
\end{figure}

In this scenario, we consider a network of seven BSs, each equipped with $L = 2$ antennas, and the fronthaul capacity per-BS is considered to be $F_n = 200$ Mbps, $\forall n \in \mathcal{N}$. Moreover, we increase the number of users from five to twenty. The maximum number of streams that can be served by a BS is set to $A_{n,\text{max}}  = 10$. Fig.~\ref{Results_user_2} shows the ESR as a function of the number of users. The ESR achieved by all schemes increases with the number of users. RS scheme 1 achieves the best ESR, thanks to its ability to address the network's interference. However, this comes with the cost of a large number of common messages, which increases the complexity of the optimization. 

In contrast, our proposed RS-CMD scheme achieves comparable gain to RS scheme 1 with a much lower number of common streams. Interestingly, in this C-RAN regime, which is a fronthaul capacity limited regime, our proposed RS-CMD scheme achieves considerable gain in both underloaded and overloaded setups. Specifically, when the number of users is eight (underloaded setup), the relative gain of RS-CMD compared to TIN is equal to 25.1\%. The gain is 22.35\% when the number of users is seventeen. This behavior indicates that RS plays an essential role in elastic management of the interference in C-RAN. As opposed to TIN, RS-CMD utilizes the common streams' extra freedom to manage the scarce fronthaul capacity wisely to enhance the performance.

To observe the effect of fronthaul capacity limit on both TIN and RS-CMD's performance, we increase the fronthaul capacity per-BS from 200 Mbps to 300 Mbps. As opposed to the previous case, the gain of RS-CMD is more pronounced as the number of users increases. Hence, for a small number of users (underloaded scenario), the interference is not substantial, as the fronthaul capacity per-BS is high enough to establish cooperation between a large number of BSs. In particular, the gain of RS-CMD compared to TIN is 5.6\% when the number of users is equal to eight, and it increases to 19.23\% in the overloaded case when the number of users is equal to seventeen.
\subsection{Ergodic Sum-rate as a function of the number of BSs}
In this scenario, we consider a C-RAN network, consisting of fifteen users; each BS is equipped with two antennas and has a fronthaul link capacity of 300 Mbps. We vary the number of BSs from four to twelve and study the achievable ESR of all the RS-schemes in addition to TIN and NOMA. Interestingly, as Fig.~\ref{Results_user_3} shows, all the RS schemes perform better than TIN. When the system is underloaded, NOMA achieves comparable performance to RS scheme 1, which deploys the largest number of common streams. However, as the number of BSs increases, the gain of all RS-schemes becomes more pronounced. Even the simple RS scheme 2 achieves a better ESR performance as compared to TIN and NOMA. The proposed RS-CMD scheme, for instance, attains a gain up to 25.2\% compared to NOMA and up to 27.9\% as compared to TIN when the number of BSs is $N = 12$. Intuitively, as the number of BSs increases, the network becomes denser, which results in improving the aggregate channel quality for each user. However, at the same time, the interference becomes larger among the users. Using the cooperative transmission scheme, C-RAN can remove part of the interference with coordinated beamforming. However, due to the limited fronthaul capacity and imperfect CSIT, the cloud can not remove all the interference. As Fig.~\ref{Results_user_3} shows, in such a network, neither decoding all the interference as in NOMA, nor ignoring the interference as in TIN gives the best performance. In contrast to the two extremes TIN and NOMA, RS techniques can manage the interference efficiently in dense networks. Moreover, our proposed RS-CMD provides a scalable and efficient transmission scheme, thus gives the best trade-off between the design complexity and good performance.
\begin{figure}
	\centering
	\includegraphics[width=0.5\linewidth]{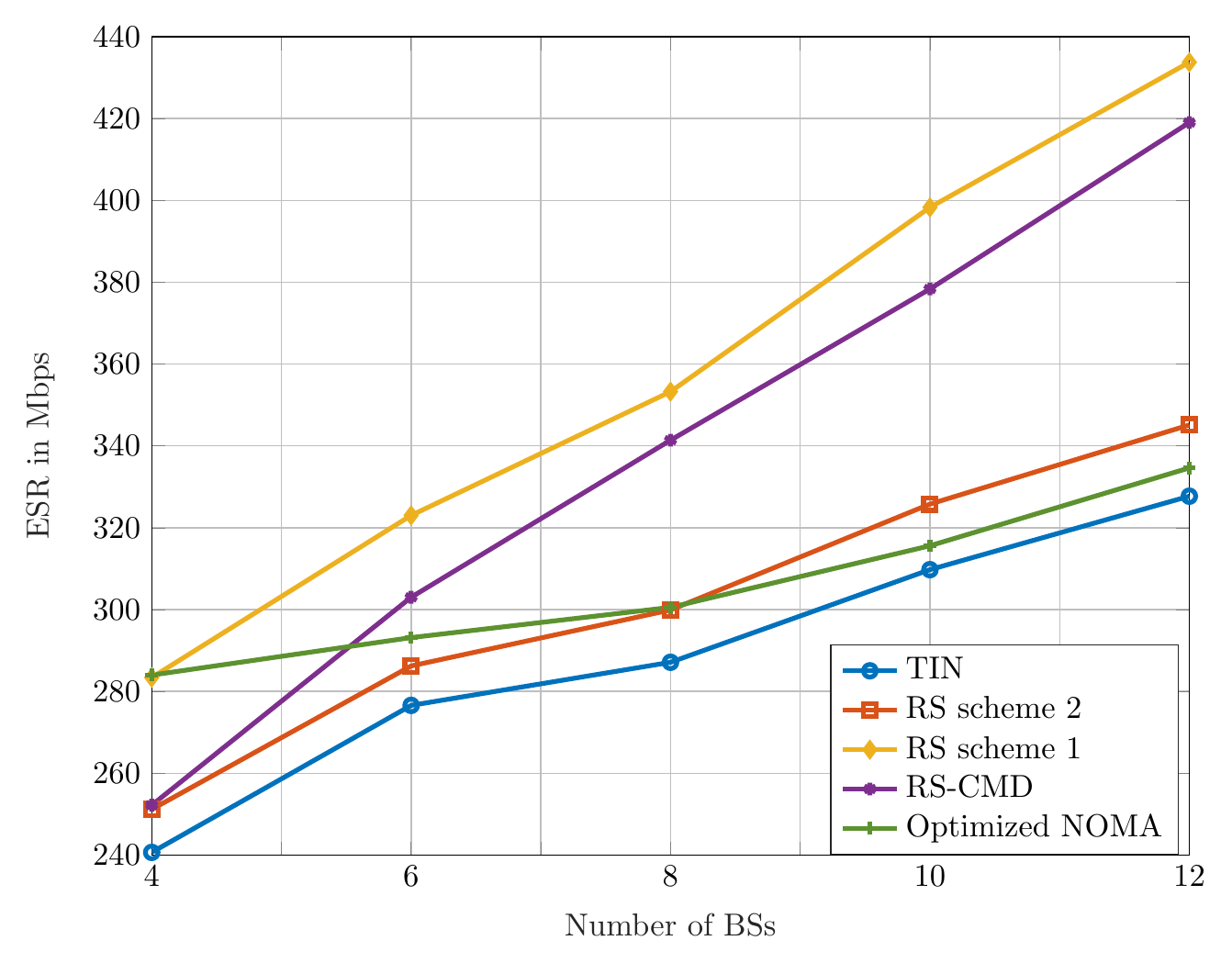}
	\caption{Ergodic sum-rate as a function of the number of base stations in the network. Fronthaul capacity per-BS is 300 Mbps.}
	\label{results_newScheme_noBS}
\vspace{-0.7cm}
\end{figure}
\section{Conclusion} \label{Con}
This paper considers employing RS-based transmission schemes in downlink C-RAN scenario with imperfect CSIT. The generalized RS and special cases thereof are investigated. Furthermore, we propose a scalable RS scheme, which is also robust to CSIT imperfections.
The numerical simulations demonstrate the significant gain of RS schemes compared to the state-of-the-art TIN and NOMA schemes, especially when the network becomes denser and when the fronthaul capacity resource is limited. Moreover, the proposed RS-CMD achieves a comparable performance to the generalized RS scheme with a significant reduction in the complexity. As for future research directions, it would be interesting to investigate the joint optimization of SCB, rate allocation and BSs' cluster design to maximize the total ESR in C-RAN.
\section{Appendix A}
First, we state the following technical assumptions to make sure the feasible set is compact and not empty.
Let us assume that the noise variance is non-zero, i.e., $\sigma > 0 $ and the transmit power is finite $P_{n}^{\text{Max}} < \infty \quad \forall n \in \mathcal{N}$. Then the SINR expressions in \eqref{eq:S1} and \eqref{eq:S2} are finite, given that the channel realizations of each user are bounded. That is, we have the following: $\gamma_{k}^p(\mathbf{w})< \infty \quad \forall k \in \mathcal{K}$, $\gamma_{k, i}^c(\mathbf{w})< \infty \quad \forall i \in \mathcal{M}_k, \forall k \in \mathcal{K}$, where the dependency of SINR expressions on the beamforming vectors is made explicit. Then, the limit of average spectral efficiency when the sample size tends to infinity exists. From ergodicity assumption of the channel distribution, and the law of large numbers the following holds \cite[Theorem 7.48]{StochSh}
\begin{align}
\label{eq:aa1}
&\sup_{\mathbf{w} \in \mathcal{W}}\left|\frac{1}{M} \log_2(1+\gamma_{k}^p(\mathbf{w})) -  \mathbb{E}_{\mathbf{h}}\left\lbrace \log_2(1+\gamma_{k}^p(\mathbf{w}))\right\rbrace\right| \to 0, \quad \text{as} \hspace{1.5mm} M \to \infty \\
\label{eq:aa2}
& \sup_{\mathbf{w} \in \mathcal{W}}\left|\frac{1}{M} \log_2(1+\gamma_{k, i}^c(\mathbf{w})) -  \mathbb{E}_{\mathbf{h}}\left\lbrace \log_2(1+\gamma_{k, i}^c(\mathbf{w}))\right\rbrace\right| \to 0, \quad \text{as} \hspace{1.5mm} M \to \infty
\end{align}
where $\mathcal{W}$ is the feasible set for the beamforming vectors, determined with $P_{n}^{\text{Max}}$ and given by the constraint \eqref{Power}. That is \eqref{eq:aa1} and \eqref{eq:aa2} indicate that the SAA estimate of the rates converges to the ergodic rate uniformly on the compact set $\mathcal{W}$ with probability one as the sample size goes to infinity. By \cite[Theorem 5.3]{StochSh}, we conclude that the set of optimal solutions of problem $\mathcal{P}_1(M)$ converges uniformly to the the set of optimal solutions of problem $\mathcal{P}_0$ with probability one as $M \to \infty$, which completes the proof.
\section{Appendix B}
The proof follows similar steps as used in \cite[Theorem 2]{razaviyayn2012linear}. First, we note that the problem $\mathcal{P}_2(M)$ is the epigraph form of the following problem
\vspace{-0.2cm}
\begin{subequations}\label{eq:Optf}
	\begin{align}
	\mathcal{Q}_{2}(M):\quad&\underset{\mathbf{w}, \mathbf{u}, \bm \rho }{\text{maximize}}\quad  \sum\nolimits_{k \in  \mathcal{K}} G_k(\mathbf{w}, \mathbf{u}, \bm \rho)  \label{eq:Objf} \\
	&\text{subject to}  \\
	&\sum_{{k \in \mathcal{K}_n^p}}\normV[\big]{\mathbf{w}_{n ,k}^{p}}_2^2  + \sum_{{k \in \mathcal{K}_n^c}}\normV[\big]{\mathbf{w}_{n ,k}^{c}}_2^2\leq P_{n}^{\text{Max}} \quad \forall n \in \mathcal{N} \label{co1}\\
	& \sum_{{k \in \mathcal{K}_n^p}} G_k^p(\mathbf{w}, \mathbf{u}, \bm \rho)  + \sum_{{k \in \mathcal{K}_n^c}} G_k^c(\mathbf{w}, \mathbf{u}, \bm \rho) \leq F_n \quad \forall n \in \mathcal{N}	\label{co2},
	\end{align}
\end{subequations}
where, $G_k(\mathbf{w}, \mathbf{u}, \bm \rho) = G_k^p(\mathbf{w}, \mathbf{u}, \bm \rho) + G_k^c(\mathbf{w}, \mathbf{u}, \bm \rho)$,\, $G_k^p(\mathbf{w}, \mathbf{u}, \bm \rho)$ and $G_k^c(\mathbf{w}, \mathbf{u}, \bm \rho)$ are defined as
\vspace{-0.2cm}
\begin{align}
& G_k^p(\mathbf{w}, \mathbf{u}, \bm \rho) = \frac{B}{M \log(2)}\sum_{m = 1}^{M}\max_{u_{k}^p(m), \rho_{k}^p(m)}\big(\log(\rho_{k}^p(m)) - \rho_{k}^p(m) e_{k}^p(m) + 1\big)\\
& G_k^c(\mathbf{w}, \mathbf{u}, \bm \rho) = \frac{B}{M \log(2)}\sum_{m = 1}^{M} \min_{i \in \mathcal{M}_k}\big( \max_{u_{k,i}^c(m), \rho_{k,i}^c(m)}\big(\log(\rho_{k,i}^c(m)) - \rho_{k,i}^c(m) e_{k,i}^c(m) + 1\big) \big)
\end{align}
Problems $P_2(M)$ and $\mathcal{Q}_{2}(M)$ are equivalent and the optimal solutions of both problems are identical \cite[Chapter 4]{Book2}. For simplicity of notations, we continue the convergence proof using the equivalent formulation $\mathcal{Q}_{2}(M)$. In each iteration $r$ of Algorithm \ref{e1}, we solve the following convex optimization problem
\vspace{-0.4cm}
\begin{subequations}\label{eq:Optf1}
	\begin{align}
	\mathcal{Q}_{2}^r:\quad&\underset{\mathbf{w}}{\text{maximize}}\quad  \sum\nolimits_{k \in  \mathcal{K}} G_k(\mathbf{w}, \mathbf{u}^{r}, \bm \rho^{r})  \label{eq:Objf1} \\
	&\text{subject to} \quad \eqref{co1}\\
	& \sum_{{k \in \mathcal{K}_n^p}} G_k^p(\mathbf{w}, \mathbf{u}^{r}, \bm \rho^{r})  + \sum_{{k \in \mathcal{K}_n^c}} G_k^c(\mathbf{w}, \mathbf{u}^{r}, \bm \rho^{r}) \leq F_n \quad \forall n \in \mathcal{N}	\label{co3}
	\end{align}
\vspace{-0.1cm}
\end{subequations}
Again, problem $P_3$ is simply the epigraph equivalent form of the problem 	$\mathcal{Q}_{2}^r$. Here, $\mathbf{u}^{r} = \Upsilon(\mathbf{w}^{r-1})$ and $\bm {\rho}^{r} = \Phi(\mathbf{w}^{r-1})$, where the mappings $\Upsilon(\cdot)$ and $\Phi(\cdot)$ are given in \eqref{e2} and \eqref{cc1}, respectively. Note that all the auxiliary coefficients in \eqref{cc2}-\eqref{e8} are already defined in terms of $\mathbf{u}^{r}$ and $\bm {\rho}^{r}$. Let us define the objective function of problem $\mathcal{Q}_2(M)$ as $Q(\mathbf{w}, \mathbf{u}, \bm \rho)$, which can be also viewed as the objective function of the problem $\mathcal{P}_2(M)$ due to the equivalence between the two optimization problem. We note that the sequence $\left\lbrace Q(\mathbf{w}^r, \mathbf{u}^r, \bm \rho^r) \right\rbrace_{r = 0}^{\infty} $ is monotonically increasing after each iteration and converges. This is because it is a concave function in each of the variables and achievable ergodic rates are bounded above for a given power constraint, let $\bar{Q}$ denotes the limit of this function. Due to the compactness of the convex feasible set defined by the constraints \eqref{co1} and \eqref{co3}, the iterates $\left\lbrace\mathbf{w}^r \right\rbrace_{r = 0}^{\infty} $ must have a cluster point, denoted as $\bar{\mathbf{w}}$. That is, it exists a subsequence $\left\lbrace\mathbf{w}^{r_1} \right\rbrace_{r_1 = J}^{\infty}$ for some $J>0$ which converges to  $\bar{\mathbf{w}}$. Since the functions $\Upsilon(\cdot)$ and $\Phi(\cdot)$ are continuous we have the following result
\vspace{-0.15cm}
\begin{equation}
\lim\limits_{r_1 \to \infty}\left\lbrace \mathbf{w}^{r_1}, \mathbf{u}^{r_1}, \bm \rho^{r_1}\right\rbrace  = \left\lbrace \bar{\mathbf{w}}, \Upsilon( \bar{\mathbf{w}}), \Phi(\bar{\mathbf{w}})\right\rbrace \triangleq \left\lbrace \bar{\mathbf{w}},  \bar{\mathbf{u}}, \bar{\bm \rho}\right\rbrace.
\vspace{-0.15cm}
\end{equation}
That is, $\left\lbrace\bar{\mathbf{u}}, \bar{\bm \rho}\right\rbrace$ is optimal when $\mathbf{w} = \bar{\mathbf{w}}$. Now, we need to prove that the beamforming vectors $\bar{\mathbf{w}}$ are optimal when $\left\lbrace{\mathbf{u}}, {\bm \rho}\right\rbrace = \left\lbrace\bar{\mathbf{u}}, \bar{\bm \rho}\right\rbrace$. To see this, we note that the $\left\lbrace\mathbf{w}^{r_1 + 1}\right\rbrace$ is given as the optimal solution of problem $\mathcal{Q}_2^{r_1}$. Hence, from the optimality of $\left\lbrace\mathbf{w}^{r_1 + 1}\right\rbrace$ and the monotonicity of the objective function we conclude the following \vspace{-0.15cm}
\begin{equation} \label{e00}
Q(\mathbf{w}^{r_1 + 1}, \mathbf{u}^{r_1 + 1}, \bm \rho^{r_1 + 1}) \geq Q(\mathbf{w}^{r_1 + 1}, \mathbf{u}^{r_1}, \bm \rho^{r_1}) \geq Q(\mathbf{w}, \mathbf{u}^{r_1}, \bm \rho^{r_1}), \quad \forall \mathbf{w}.
\vspace{-0.15cm}
\end{equation}
By taking the limits of both sides of equation \eqref{e00}, we get \vspace{-0.15cm}
\begin{equation} \label{e01}
\bar{Q} = Q(\bar{\mathbf{w}}, \bar{\mathbf{u}}, \bar{\bm \rho}) \geq Q(\mathbf{w}, \bar{\mathbf{u}}, \bar{\bm \rho}), \quad \forall \mathbf{w}.
\vspace{-0.15cm}
\end{equation}
Thus, $\bar{\mathbf{w}}$ must be the optimal solution to the problem $\mathcal{Q}_2(M)$ when $\left\lbrace{\mathbf{u}}, {\bm \rho}\right\rbrace = \left\lbrace\bar{\mathbf{u}}, \bar{\bm \rho}\right\rbrace$, and we already have shown that $\left\lbrace\bar{\mathbf{u}}, \bar{\bm \rho}\right\rbrace$ is the optimal solution to problem $\mathcal{Q}_2^(M)$ when $\mathbf{w} = \bar{\mathbf{w}}$. Based on these observations we can easily show that $\left\lbrace \bar{\mathbf{w}}, \bar{\mathbf{u}}, \bar{\bm \rho}\right\rbrace$ is a KKT solution to the optimization problem $\mathcal{Q}_2(M)$ by checking the KKT conditions. To this end, we have shown that any cluster point of the iterates generated by Algorithm \ref{e1} converges to a KKT solution of the optimization problem $\mathcal{Q}_2(M)$. Although the KKT points are not necessarily unique, the distance between any sequence $\left\lbrace\mathbf{w}^r, \mathbf{u}^r, \bm \rho^r \right\rbrace_{r = 0}^{\infty}$  and the KKT solution set goes to zero in the limit. Due to the equivalence between $\mathcal{Q}_2(M)$ and $P_2(M)$, we conclude that the tuple  $ \left\lbrace \bar{\mathbf{w}}, \bar{\mathbf{u}}, \bar{\bm \rho},\bar{\mathbf{R}}\right\rbrace$ is also a KKT solution to problem $P_2(M)$ where $\bar{\mathbf{R}} = \left\lbrace \bar{{R}}_k^p, \bar{{R}}_k^c|\hspace{1mm} \forall k \in \mathcal{K}\right\rbrace $ and the following holds in the limit
\begin{equation}
Q(\bar{\mathbf{w}}, \bar{\mathbf{u}}, \bar{\bm \rho}) = \sum_{{k \in \mathcal{K}}}\left( \bar{{R}}_k^p +  \bar{{R}}_k^c \right).
\end{equation}
This completes the proof.
\bibliographystyle{IEEEtran}
\bibliography{bibliography}
\balance
\end{document}